\documentclass[preprint,11pt]{elsarticle}
\usepackage{graphicx}
\usepackage{verbatim}
\usepackage{epsfig}
\usepackage{amsmath}
\usepackage{amsfonts}
\usepackage{amssymb}
\usepackage{mathrsfs}
\usepackage{rotating}
\usepackage{mathtools}
\usepackage{multirow}
\usepackage{amsthm}
\usepackage{multicol}
\usepackage{subcaption}
\usepackage{algorithm}
\usepackage{algorithmic}
\usepackage{enumerate}
\usepackage{bm}
\usepackage{hyperref}
\def\be{\begin{equation}}
\def\ee{\end{equation}}
\def\bq{\begin{eqnarray}}
\def\eq{\end{eqnarray}}

\newtheorem{theorem}{Theorem}[section]

\newtheorem{remark}[theorem]{Remark}

\makeatletter
\def\ps@pprintTitle{%
  \let\@oddhead\@empty
  \let\@evenhead\@empty
  \def\@oddfoot{\reset@font\hfil\thepage\hfil}
  \let\@evenfoot\@oddfoot
}
\makeatother
\usepackage{amssymb}

\begin{document}
\begin{frontmatter}
	
\title{Neural Network for valuing Bitcoin options under jump-diffusion and market sentiment model}
\author{Edson Pindza\footnotemark[1]; Jules Clement Mba\footnotemark[2];  Sutene Mwambi\footnotemark[2]; Nneka Umeorah\footnotemark[3]}
\address{\footnotemark[1]Tshwane University of Technology; Department of Mathematics and Statistics; 175 Nelson Mandela Drive OR Private Bag X680; Pretoria 0001; South Africa [edsonpindza@gmail.com]\\
\footnotemark[2]University of Johannesburg; School of Economics, College of Business and Economics; P. O. Box 524, Auckland Park 2006; South Africa [jmba@uj.ac.za \& sutenem@uj.ac.za]\\
\footnotemark[3]Cardiff University; School of Mathematics; Cardiff CF24 4AG; United Kingdom [umeorahn@cardiff.ac.uk]\\

\textbf{Correponding author}: Nneka Umeorah [umeorahn@cardiff.ac.uk] }

\begin{abstract}
Cryptocurrencies and Bitcoin, in particular, are prone to wild swings resulting in frequent jumps in prices, making them historically popular for traders to speculate. A better understanding of these fluctuations can greatly benefit crypto investors by allowing them to make informed decisions. It is claimed in recent literature that Bitcoin price is influenced by sentiment about the Bitcoin system. Transaction, as well as the popularity, have shown positive evidence as potential drivers of Bitcoin price. This study considers a bivariate jump-diffusion model to describe Bitcoin price dynamics and the number of Google searches affecting the price, representing a sentiment indicator. We obtain a closed formula for the Bitcoin price and derive the Black-Scholes equation for Bitcoin options. We first solve the corresponding Bitcoin option partial differential equation for the pricing process by introducing artificial neural networks and incorporating multi-layer perceptron techniques. The prediction performance and the model validation using various high-volatile stocks were assessed.\\

\end{abstract}

%\begin{keyword}
%L\'{e}vy process \sep Memory dependence  \sep Bitcoin\sep Forecasting.\\
%\emph{JEL}:  C15 \sep C53 \sep  G17.
%\end{keyword}
\end{frontmatter}

%\footnotemark [1]{Correspondence to: Jules Clement Mba,
%E-mail: jmba@uj.ac.za}

\section{Introduction}
\cite{sene2021pricing},\cite{olivares2020pricing},\cite{chen2021detecting}, \cite{grohs2018proof}
\noindent Bitcoin, a decentralized network-based digital currency and payment system, is a special type of cryptocurrency developed in 2009 \cite{nakamoto2008re} by a person or a group of persons known under the name of Satoshi Nakamoto. The soar in bitcoin appreciation has been accompanied by high uncertainty and volatility, which surrounds future prices, and this has attracted rapidly increasing research into this digital asset. Policymakers globally are concerned whether bitcoin is decentralized and unregulated and whether it could be a bubble \cite{cheah2015speculative} which threatens the stability of a given financial system. Regardless of the speculation, traders are still interested in the capacity of Bitcoin to generate more returns and the use of bitcoin derivatives as a risk management and diversification tool. Blockchain is the underlying technology that powers Bitcoin by recording transactions in a distributive manner, removing alteration and censorship. Following the success of Bitcoin and its growing community, many other alternatives to Bitcoin have emerged. There are more than 5000 tradable cryptocurrencies \footnote{\url{https://coinmarketcap.com/}} with a total market capitalization of USD 248 billion at the time of this writing (see \cite{yermack2015bitcoin,yermack2017corporate} for further background on Bitcoin and its technology).\\ 

On the one hand, the cryptocurrency market is known to be highly volatile (\cite{dwyer2015economics,katsiampa2017volatility}) due to its sensibility to new information, whether fundamental or speculative \cite{cheah2015speculative} since it does not rely on the stabilizing policy of a central bank. On the other hand, the relative illiquidity of the market with no official market makers makes it fundamentally fragile to large trading volumes and market imperfections and thus more prone to large swings than other traded assets, see \cite{scaillet2017high}. This concept results in frequent jumps of larger amplitude than what a continuous diffusion process can explain. Due to its local Markov property, i.e., the asset price changes only by a small amount in a short interval of time. When analyzing cryptocurrency data, it is interesting to consider processes that allow for random fluctuations that have more than a marginal effect on the cryptocurrency's price. Such a stochastic process that enables us to incorporate this type of effect is the jump process. This process allows the random fluctuations of the asset price to have two components, one consisting of the usual increments of a Wiener process; the second provides for ``large" jumps in the asset price from time to time. Shortly after the development of the Black–Scholes option valuation formula, Merton \cite{merton1976option} developed a jump-diffusion model to account for the excess kurtosis and negative skewness observed in log-return data (see, \cite{Matsuda2004IntroductionTM}). This jump process is appended to the Black–Scholes geometric stochastic differential equation.\\

Several studies have used the daily bitcoin data to document the impact of jump as a crucial feature of the cryptocurrency dynamics. Chaim and Fry \citep{chaim2018volatility} observed that jumps associated with bitcoin volatility are permanent, whereas the jumps to mean returns are said to have contemporaneous effects. The latter equally capture large price bubbles, mainly negative, and are often associated with hacks and unsuccessful fork attempts in the cryptocurrency markets. Hillard and Ngo \citep{hilliard2022bitcoin} further investigated the characteristics of bitcoin prices and derived a model which incorporates both jumps and stochastic convenience yield. The result was tested with data obtained from the Deribit exchange, and they observed that modelling bitcoin prices as a jump-diffusion model outperformed the classical Black-Scholes models. Philippas \emph{et al.} \citep{philippas2019media} observed that during periods of high uncertainty, some informative signals, which are proxied by Google search volume and Twitter tweets, have a partial influence on Bitcoin prices and price jumps.\\

In recent years, the focus has been on identifying the main drivers of Bitcoin price evolution in time. Many researchers have assigned the high volatility in Bitcoin prices to the sentiment and popularity of the Bitcoin system. Though these are not directly observable, they may be considered as indicators from the transaction volumes or the number of Google searches or Wikipedia requests about the topic, see for example, Kristoufek \cite{kristoufek2013bitcoin,kristoufek2015main}, Kim et al. \cite{kim2015virtual} and \cite{bukovina2016sentiment}. Authors in \cite{bukovina2016sentiment} use a Bitcoin sentiment measure from Sentdex.com and develop a discrete-time model to show that excessive confidence in the system may boost a bubble in the Bitcoin system. These sentiment-based data were collected through Natural Language Processing techniques to identify a string of words conveying positive, neutral or negative sentiment on Bitcoin. Furthermore, the authors in \cite{cretarola2017sentiment} introduce a bivariate model in continuous time to describe both the dynamics of the Bitcoin sentiment indicator and the corresponding Bitcoin price. By fitting their bivariate model to market data, they consider both the volume and the number of Google searches as proxies for the sentiment factor.\\

While traditional financial theory relies on assumptions of market efficiency, normally distributed returns, and no-arbitrage, emerging research suggests cryptocurrency markets exhibit different characteristics. These markets appear prone to inefficiencies, fat-tailed non-normal distributions, and frequent arbitrage opportunities according to  \citet{kabavsinskas2021key}. For example, the extreme volatility and relative illiquidity of cryptocurrencies can lead to dislocations between markets where arbitrageurs can profit. Additionally, the return distributions tend to exhibit higher kurtosis and skewness than normal distributions. However, despite violating some traditional assumptions, jump-diffusion models can still provide a useful starting point for modelling cryptocurrency dynamics. The jumps can account for extreme price fluctuations beyond what continuous diffusion alone predicts. While the model may require modifications over time as more data becomes available, it captures key features like volatility clustering and significant outliers. The sentiment indicator variable also represents an initial attempt to incorporate a behavioural factor affecting prices.\\

Our technique incorporates the one similar to Cretarola \emph{et al.} \cite{cretarola2017sentiment}, who developed a continuous bivariate model that described the bitcoin price dynamics as one factor, and a sentiment indicator as the second factor. We further added a jump-diffusion component to the SDE with the aim of capturing the occurrence of rare or extreme events in the bitcoin price return. Furthermore, we introduce the artificial neural network and propose a trial solution that solves the associated Black-Scholes partial differential equation (PDE) for the bitcoin call options with European features. This concept is equally different from the univariate jump-diffusion model used, for example, by Chen and Huang (2021). We further implemented the number of Google searches as a Bitcoin sentiment indicator in this paper. This choice is due to their unique transparency in contrast to other social media-driven measures, and they have the tendency to gauge behaviour instead of searching for it. Therefore, using search-based data as sentiment indices has the potential to reveal the underlying beliefs of populations directly.\\

In this paper, we develop an initial modelling framework using a bivariate jump-diffusion model and sentiment indicator. This provides a foundation for pricing and derivatives valuation in cryptocurrency markets. We acknowledge the model's limitations per the evolving understanding of these new markets. As future research expands knowledge of distributional properties, market microstructure issues, and other intricacies, the modelling approach can be enhanced. Nonetheless, our proposed model offers an initial step toward financial engineering in the cryptocurrency space. To this effect, the significant contributions of this paper are highlighted as follows:

\begin{itemize}
\item We present a bivariate jump-diffusion model to describe the dynamics of the Bitcoin sentiment indicator, which consists of volumes or Google searches and the corresponding Bitcoin price.
\item We derive a closed-form formula for the Bitcoin price and the corresponding Black-Scholes equation for Bitcoin options valuation.
\item The corresponding bitcoin option PDE is solved using the artificial neural network, and the proposed model was validated using data from highly volatile stocks.
\end{itemize}
 The rest of the paper is organized as follows: Section 2 introduces the methodology and highlights the strengths and the limitations of the proposed model. Section 3 describes the concept of the artificial neural network, as well as its applications in solving the option pricing differential equations, Section 4 discusses the numerical implementation findings, and the last section concludes the work.

\section{Methodology}

Jump-diffusion models are continuous-time stochastic processes introduced in quantitative finance by Merton \cite{merton1973theory}, extending the work on option pricing by Black and Scholes \cite{black1973pricing}. These models reproduce stylized facts observed in asset price dynamics, such as mean-reversion and jumps. 
%\section{Description and Interpretation of Jump Process}
There are two ways to expound the description of jump processes.
\begin{enumerate}
	\item One is to describe the jump in \textbf{ absolute terms} (this is useful if we are focusing on
	prices or interest rates),
	\item The other is to describe the jump in \textbf{ proportional terms} (this
	is more useful if our focus is on returns, as in this study).
\end{enumerate}

In addition to jump-diffusion models, Levy processes appear as an alternative for capturing large deviations in asset prices. Levy processes allow sample paths with frequent discontinuities, enabling them to generate heavy-tailed distributions. Some key examples include variance gamma processes, normal inverse Gaussian processes, and generalized hyperbolic processes. \citep{watanabe2006} demonstrated that Bitcoin returns exhibit heavy tails and proposed using a variance gamma process to model the price dynamics. \citep{kim2017bitcoin} found evidence that variance gamma processes fit cryptocurrency log returns compared to classical diffusion. Levy processes have also been employed to model stochastic volatility in Bitcoin prices \citep{caporale2018persistence}. Overall, Levy processes provide an alternative class of models with more flexibility to account for extreme deviations. They have shown promise in fitting the empirical distributional properties of cryptocurrency returns. As this work focuses initially on extending jump-diffusion models, exploring Levy processes represents a worthwhile direction for future research. However, their ability to capture heavy tails and discontinuities suggests that  Levy processes could serve as a valuable modelling technique in this domain.\\

Consider a probability space $(\Omega,{\cal F},{\mathbf{P}})$ endowed with a filtration $\mathbb{F}=\{{\cal F}_t,\ t\geq 0\}$ that satisfies the usual conditions of right-continuity and completeness, where ${\cal F}_t=\sigma\big(W_s,N_s;\ 0\leq s\leq t\big)$: $W_t$ is a standard Brownian motion, and $N_t$ denotes a counting process that can be represented as $N_t=\sum_{i}1_{\{T_i\leq t\}}$ with a sequence of stopping times $0<T_1<T_2<\cdots<T$ whose number is finite with probability one.\\ 

Let $S=\{S_t,\ t\geq 0\}$ be the price process of Bitcoin, and let $y_t$ be the absolute jump size, where we assume that the bitcoin price jumps from $S_t$ to $y_tS_t$ in a small time interval $\mathbb{d}t$. The relative price jump size, $J_t= y_t-1$, is the percentage change in the bitcoin price influenced by the jump. Using the Merton jump-diffusion model, the absolute price jump size is a non-negative random variable which is log-normally distributed, i.e $\ln (y_t) \sim i.i.d \; \mathcal{N}(\mu, \delta^2)$ \cite{matsuda2004introduction}. Thus, we further assume that the Bitcoin price dynamics are described by the following jump-diffusion stochastic differential equation
\begin{align}\label{mjd}
dS_t= S_t P_{t-\tau}(\mu_d-\lambda k)dt + \sigma_d\sqrt{P_{t-\tau}}S_t dW_t+ S_t P_{t-\tau}(y_t-1)dN_t,\qquad S_0=s_0\in \mathbb{R}_+ 
\end{align}
where $\mu_d$ is the diffusion mean, $\sigma_d$ the diffusion volatility, $\lambda$ the jump intensity rate and $\mathbb{E}[J_t]=\mathrm{e}^{\mu + \frac{\delta^2}{2}}-1=k$, the expected proportional jump size. The one-dimensional Poisson process is discontinuous with a constant jump $\lambda$ and is given by
\[\mathrm{d}N_t = N_{t+dt}-N_t=\begin{cases}
0 \; \text{with probability}\; 1-\lambda \mathrm{d}t,\\
1 \; \text{with probability}\; \lambda \mathrm{d}t
\end{cases}  \,.\]
Also, the exogenous process $P=\{P_t,\ t\geq 0\}$ is a stochastic factor representing the sentiment index in the Bitcoin market, satisfying
\begin{align}\label{st}
	dP_t= P_t\mu_p dt + \sigma_pP_t dZ_t,\qquad P_t=\phi(t),\ t\in [-L,0].
\end{align}
where $\mu_p\in \mathbb{R}-\{0\}$, $\sigma_p\in \mathbb{R}_+$, $L \in \mathbb{R}_+$, $Z=\{Z_t,\ t\geq 0\}$ is a standard $\mathbb{F}$-Brownian motion on $(\Omega,{\cal F},{\mathbf{P}})$, which is $\mathbf{P}$-independent of $W$, and $\phi:[-L,0]\to[0,+\infty)$ is a continuous (deterministic) initial function. The non-negative property of the function $\phi$ corresponds to the requirement that the minimum level for sentiment is zero. The delay variable $P_{t-\tau}$ accounts for the effect of investor sentiment on volatility, and this is supported by findings that factors like public attention and transaction volume influence cryptocurrency volatility (e.g. \citep{bukovina2016sentiment} and \citep{cretarola2017sentiment}). This variable affects the instantaneous variance of the Bitcoin price modulated by $\sigma_d$. The sentiment index may be the volume/the number of Bitcoin transactions or the number of internet searches within a fixed time period. The solution to equation (\ref{st}) exists in its closed form \cite{black1973pricing}, and $P_t$ is lognormally distributed for each $t > 0$. \\

The system given by equations \ref{mjd} and \ref{st} is well-defined in $\mathbb{R}_+$ as stated in the following theorem

\begin{theorem}
	In the market model described previously, the following holds:
\begin{enumerate}
	\item the bivariate stochastic delayed differential equation
	\begin{equation}
	\begin{cases}
		dS_t= S_t P_{t-\tau}(\mu_d-\lambda k)dt + \sigma_d\sqrt{P_{t-\tau}}S_t dW_t+ S_t P_{t-\tau}(y_t-1)dN_t,\qquad S_0=s_0\in \mathbb{R}_+ \\
		dP_t= P_t\mu_p dt + \sigma_pP_t dZ_t,\qquad P_t=\phi(t),\ t\in [-L,0].	
	\end{cases}
	\end{equation}
has a continuous, $\mathbb{F}$-adapted, unique strong solution $(S,P)=\{(S_t,P_t),\ t\geq 0\}$ given by
\begin{equation}\label{sol1}
S_t=S_0\exp\bigg(\big((\mu_d-\lambda k)-\frac{\sigma_d^2}{2}\big)\int_{0}^{t}P_{u-\tau}du+\sigma_d\int_{0}^{t}\sqrt{P_{u-\tau}}dW_u+\int_{0}^{t}\ln(1+P_{u-\tau}(y_u-1))dN_u\bigg)	
\end{equation}
\begin{equation}\label{sol2}
P_t=\phi(0)\exp\bigg(\big(\mu_d-\frac{\sigma_d^2}{2}\big)t+\sigma_dZ_t\bigg) ,\qquad t\geq 0	
\end{equation} 
\end{enumerate}	
\end{theorem}
The solution (\ref{sol1}) can be obtained by applying the Ito's formula for jump-diffusion processes taken from \cite[Proposition 8.14, p.275]{tankov2003financial} to $\ln(S_t)$ and assuming that the process $S_t$ is c\'{a}gl\'{a}d. 
\begin{multline*}
	\ln(S_t)=\ln(S_0)+\int_{0}^{t}P_{u-\tau}(\mu_d-\lambda k)du+\dfrac{1}{2}\int_{0}^{t}\sigma_d^2P_{u-\tau}du+\int_{0}^{t}\sigma_d\sqrt{P_{u-\tau}}dW_u+\\
\int_{0}^{t}\ln(S_{u^-}+S_{u}P_{u-\tau}(y_u-1))-\ln(S_{u^-})\ dN_u \,.
\end{multline*}

\subsection{Strengths and limitations of the proposed model}

The bivariate jump-diffusion model provides an essential first step in applying financial engineering techniques to this new domain. By capturing stochastic volatility, discrete jumps, and a sentiment indicator, it incorporates several salient features of cryptocurrencies. The preliminary pricing and derivatives valuation framework can be built upon as the market evolves. While the assumptions require validation as more data emerges, the model offers valuable inputs for investment analysis at present. We acknowledge the need to refine techniques as cryptocurrency finance matures into its own field. This paper aims to provide the initial modelling foundation, which can be improved incrementally as research progresses. Future enhancements may include more flexible return distributions, multifactor models to capture additional risks, regime-switching models to reflect market phases, and calibration of the sentiment factor from various data sources.\\

In addition, the current proposed model offers a useful starting point, we recognize this initial framework may require modifications over time to align with market realities. As more cryptocurrency data becomes available, the distributional properties and other intricacies of these markets will become clearer. If the actual return distributions exhibit higher kurtosis than the normal distribution assumed, the model may need to be adjusted accordingly. Furthermore, if inefficiencies and arbitrage opportunities remain prevalent, the dynamics may diverge from a pure jump-diffusion process. As researchers expand their knowledge of the microstructure, behavioural components, and risks in cryptocurrency markets, our modelling approach can be enhanced. In summary, we emphasize this work represents a starting point for modelling cryptocurrencies and enabling financial applications. The proposed bivariate jump-diffusion model and sentiment indicator capture essential dynamics but may require adjustments as knowledge develops. We welcome future research to build on these initial techniques for pricing and valuation of cryptocurrency-denominated assets.

\subsection{Applications}

We now derive the Black-Scholes PDE by a hedging argument. Consider the stock price process described by the equations \ref{mjd} and \ref{st}. Let $B(t)$ be the risk-free asset process:
\[B(t)=e^{rt},\ \ \ t\in[0,T]\] 
where $r>0$. Let $V_t=V(t,S_t)$ be the price of the derivative that cannot be exercised before maturity. Applying Ito's formula,\\
\begin{multline*}
	dV(t,S_t)=\bigg(\dfrac{\partial V}{\partial t}(t,S_t)+(\mu_d-\lambda k)S_tP_{t-\tau}\dfrac{\partial V}{\partial S}(t,S_t)+\dfrac{\sigma^2_d}{2}P_{t-\tau}S^2_t\dfrac{\partial^2 V}{\partial S^2}(t,S_t)\bigg)dt+\\ \sigma_d\sqrt{P_{t-\tau}}S_t\dfrac{\partial V}{\partial S}(t,S_t)dW_t+\Delta V(t,S_t)dN_t
\end{multline*}
We set up a self-financing portfolio $X$ that is comprised of one option and $\delta$ amount of the underlying stock, such that the portfolio is riskless, that is insensitive to changes in the price of the security. So, at time $t$, the value of the portfolio is $X_t=V_t+\delta S_t$. The self-financing assumption implies that 
\begin{equation}\label{sfp}
	dX_t=dV_t+\delta dS_t
\end{equation}
Substituting the $dV_t$ and $dS_t$ into the Equation (\ref{sfp}), we have\\
\begin{multline*}
	dX_t=\biggl[\bigg(\dfrac{\partial V}{\partial t}(t,S_t)+(\mu_d-\lambda k)S_tP_{t-\tau}\dfrac{\partial V}{\partial S}(t,S_t)+\dfrac{\sigma^2_d}{2}P_{t-\tau}S^2_t\dfrac{\partial^2 V}{\partial S^2}(t,S_t)\bigg)dt+\\ \sigma_d\sqrt{P_{t-\tau}}S_t\dfrac{\partial V}{\partial S}(t,S_t)dW_t+\Delta V(t,S_t)dN_t\biggr]+\delta\biggl[S_t P_{t-\tau}(\mu_d-\lambda k)dt + \sigma_d\sqrt{P_{t-\tau}}S_t dW_t+ S_t P_{t-\tau}(y_t-1)dN_t\biggr]
\end{multline*}
So,
\begin{multline}\label{sfp2}
	dX_t=\bigg(\dfrac{\partial V}{\partial t}(t,S_t)+(\mu_d-\lambda k)S_tP_{t-\tau}\bigg(1+\dfrac{\partial V}{\partial S}(t,S_t)\bigg)+\dfrac{\sigma^2_d}{2}P_{t-\tau}S^2_t\dfrac{\partial^2 V}{\partial S^2}(t,S_t)\bigg)dt+\\ \sigma_d\sqrt{P_{t-\tau}}S_t\bigg(\delta +\dfrac{\partial V}{\partial S}(t,S_t)\bigg)dW_t+\bigg(\delta S_t P_{t-\tau}(y_t-1)+\Delta V(t,S_t)\bigg)dN_t
\end{multline}
On one hand, this portfolio should be riskless and earn a risk-free rate. To be riskless, the second and the third terms involving, respectively, the Brownian motion $dW_t$ and the Poisson process $dN_t$ must be zero. That is\\
\[\delta=-\dfrac{\partial V}{\partial S}(t,S_t)\ \text{and}\ \Delta V(t,S_t)=-\delta S_t P_{t-\tau}(y_t-1)\]
Hence,
\[\Delta V(t,S_t)=S_t P_{t-\tau}(y_t-1)\dfrac{\partial V}{\partial S}(t,S_t)\]
Substituting for $\delta$ and $\Delta V(t,S_t)$ in Equation (\ref{sfp2}) implies that the portfolio follows the process
\begin{equation}\label{sfp3}
dX_t=\bigg(\dfrac{\partial V}{\partial t}(t,S_t)+(\mu_d-\lambda k)S_tP_{t-\tau}\bigg(1+\dfrac{\partial V}{\partial S}(t,S_t)\bigg)+\dfrac{\sigma^2_d}{2}P_{t-\tau}S^2_t\dfrac{\partial^2 V}{\partial S^2}(t,S_t)\bigg)dt	
\end{equation}
On the other hand, the portfolio must earn the risk-free rate, that is 
\[dX_t=rX_tdt\]
\[\bigg(\dfrac{\partial V_t}{\partial t}+(\mu_d-\lambda k)S_tP_{t-\tau}\bigg(1+\dfrac{\partial V_t}{\partial S}\bigg)+\dfrac{\sigma^2_d}{2}P_{t-\tau}S^2_t\dfrac{\partial^2 V_t}{\partial S^2}\bigg)dt=r\bigg(V_t-\dfrac{\partial V_t}{\partial S} S_t\bigg)dt\]
Multiplying both sides by $\dfrac{1}{dt}$ and re-arranging yields the following Black-Scholes PDE for an option $V$\\
\begin{equation}\label{bs}
\dfrac{\partial V_t}{\partial t}+\dfrac{\sigma^2_d}{2}P_{t-\tau}S^2_t\dfrac{\partial^2 V_t}{\partial S^2}+rS_t\dfrac{\partial V_t}{\partial S} -rV_t=-(\mu_d-\lambda k)S_tP_{t-\tau}\bigg(1+\dfrac{\partial V_t}{\partial S}\bigg)	
\end{equation}

This extended Black-Scholes PDE poses challenges for analytical solutions due to the unbounded upside of the European call payoff. Therefore, numerical methods are often used to approximate the solution. Two standard techniques are finite difference methods and Monte Carlo simulation.
Finite difference methods involve discretizing the price domain and time dimensions and replacing the derivatives with finite difference approximations. This transforms the PDE into a system of algebraic equations that can be solved recursively. Various finite difference schemes can be employed, such as explicit, implicit, and Crank-Nicolson.
Monte Carlo simulation generates sample paths of the underlying price using the assumed stochastic process. The discounted payoffs from each sample path are averaged to estimate the option price. Variance reduction techniques like antithetic sampling and control variates can improve efficiency. Both methods have tradeoffs regarding accuracy, speed, and implementation complexity. Our proposed neural network approach serves as an alternative way to numerically solve the pricing PDE without discretizing the domain. This offers benefits in terms of generalization and scaling.

\section{Neural Network Methodology}

The neural network (NN) algorithm is used for solving problems relating to dimensionality reduction \citep{sarveniazi2014actual, teli2007dimensionality}, visualization of data \citep{marghescu2007multidimensional}, clustering \citep{du2010clustering, xu2005survey} and classification issues \citep{nazemi2015neural, o2001combining}. It is also used in regression \citep{bataineh2017neural, jiang2022efficient, setiono2004approach} in solving regression problems since they do not require prior knowledge about the distribution of the data. The NN algorithm often has the tendency to predict with higher accuracy than other techniques due to the NN's capability to fit any continuous functions \citep{hornik1991approximation}. Mathematically, the NN can be referred to as a directed graph having neurons as vertices and links as edges. Different forms of NN depend on the connectivity of their neurons, and the simplest one is the Multi-layer perceptron, also known as the feedforward NN. Basically, a NN can have only one input and one output layer in its simplest form, and this can be written as:
\[y_k= \Phi \left(b_k + \sum_{i=1}^m w_{i,k}\bm{x}_i \right)  \,,\] 

where $m$ is the number of input variables, $\Phi$ is the activation function, $w_{i,k}$ is the weight of the input layer $i$ with respect to the output $k$,  $b$ is the bias, and the input vector $\bm{x}$ is connected to the output $k$ (denoting the $k$th neuron in the output layer) through a biased weighted sum. In the presence of hidden layers positioned between the input layers and the output layers, the output can be written as:
\[y_p= \Phi_{out} \left[b_p^{(2)} + \sum_{k=1}^{m_2} w_{k,p}^{(2)} \times \Phi \left(b_k^{(1)} +\sum_{i=1}^{m_1} w_{i,k}^{(1)} \ \bm{x}_i \right)\right]  \,,\]
where $m_1$ is the number of input variables, $m_2$ is the number of nodes in the hidden layer, $\Phi: \mathbb{R} \rightarrow \mathbb{R}$ refers to non-linear activation functions for each of the layers in the brackets, and $\Phi_{out}$ is the possibly new output function.\\

Recently, the NN has become an indispensable tool for learning the solutions of differential equations, and this section will utilize the potential of the NN in solving PDE-related problems \citep{aarts2001neural, dissanayake1994neural, hussian2015numerical, umeorah2022approximation}. Also, the NN has been implemented in solving equations without analytical solutions. For instance, \citet{nwankwo2023deep} considered the solution of the American options PDE by incorporating the Landau transformation and solving the corresponding transformed function via a neural network approach. In many scientific and industrial contexts, there is a need to solve parametric PDEs using NN techniques and \citet{khoo2021solving} developed such a technique which solves PDE with inhomogeneous coefficient fields. For high-dimensional parametric PDE, \citep{glau2022deep} analyzed the deep parametric PDE method to solve high-dimensional parametric PDEs with much emphasis on financial applications.\\

In the following, we describe the Bitcoin options with the corresponding Black-Scholes pricing PDE:

\begin{equation}\label{a}
\frac{\partial V}{\partial t} + \frac{\sigma^2 S^2}{2}P_{t-\tau}\frac{\partial^2 V}{\partial S^2} + (r + (\mu-\lambda K)P_{t-\tau})S\frac{\partial V}{\partial S} -rV + (\mu-\lambda K)S P_{t-\tau}=0
\end{equation}

The above equation which is a non-homogeneous PDE can be re-written as 
\begin{equation}\label{b}
\frac{\partial V}{\partial t} + \frac{\sigma_*^2 S^2}{2} \frac{\partial^2 V}{\partial S^2} + \eta S\frac{\partial V}{\partial S} -rV = \beta(S)\,,
\end{equation}

where
$\sigma_*^2=\sigma^2 P_{t-\tau}, \;\; \eta= r + (\mu-\lambda K)P_{t-\tau},\;\; \beta(S)=-(\mu-\lambda K)S P_{t-\tau}$ and \\
$P_t = \phi(0) \exp\{(\mu_p-\frac{\sigma_p^2}{2})t + \sigma_p Z_t \}$.\\

Equation (\ref{b}) can be written in its operator format. Let $\Omega=S_{max}$ and $\mathcal{A}$ be the infinitesimal operator for the stochastic process defined by

\begin{equation}
\mathcal{A}= S \eta (t,s) \frac{\partial}{\partial S} + \frac{S^2 \sigma_*^2(t,S)}{2}\frac{\partial^2}{\partial S^2}\,,
\end{equation}

with both the boundary and the terminal value problem written as
\begin{align}\label{c}
(\partial_t + \mathcal{A}-r)V(t,S) &=\beta(t,S), \qquad \qquad t,S \in [0,T] \times \Omega \nonumber \\\nonumber
V(T,S)&=g_1(T,S)=\text{Payoff}, \qquad \qquad t,S \in [0,T] \times \partial \Omega \\\nonumber
V(t,S)&=g_2(t,S)=S_t-K\mathrm{e}^{-r(T-t)}, \qquad \qquad S \in \Omega\\
V(t,0)&=V_0(t)=0, \qquad \qquad t \in [0,T]\,.
\end{align}

Standard theorems guarantee a classical smooth solution exists for the pricing PDE (3.13) given the assumed dynamics. The continuity and linear growth conditions on the coefficient functions ensure they satisfy Lipschitz continuity. Contingent claims theory shows that Lipschitz continuity is sufficient for existence and uniqueness under the stochastic process specifications. In particular, the smooth past price dependence in the diffusion coefficient allows the application results for delayed Black-Scholes equations. Therefore, the Cauchy problem is well-posed under the model assumptions, and a smooth pricing solution is guaranteed to exist based on the theorems. This provides the theoretical foundation for using a neural network to approximate the price function numerically. Thus, employing the ANN to solve the PDE, we introduce an approximating function $\zeta(t,S:\bm{\theta})$ with parameter set $\bm{\theta}$. The loss or cost function associated with this training is measured by how well the approximation function satisfies the differential operator, boundary conditions and the terminal condition of the option pricing PDE. These are given respectively as
\begin{itemize}
\item Differential operator $\left\Vert \partial_t + \mathcal{A}-r)\zeta(t,S:\bm{\theta}) \right\Vert^2_{Y_1}$, where $Y_1=[0,T] \times \Omega, v_1$.
\item Terminal condition $\left\Vert \zeta(T,S:\bm{\theta})-g_1(T,S) \right\Vert^2_{Y_2}$, where $Y_2=\Omega,v_2$.
\item Boundary condition $\left\Vert \zeta(t,0:\bm{\theta})-V_0(t) \right\Vert^2_{Y_3}$, where $Y_3=[0,T],v_3$.
\item Boundary condition $\left\Vert \zeta(t,S:\bm{\theta})-g_2(t,S) \right\Vert^2_{Y_4}$, where $Y_4=[0,T] \times \partial\Omega,v_4$. 
\end{itemize}

In all these four terms above, the observed error is measured in the Hilbert space $L^2$-norm, meaning that 
$\left\Vert \lambda(x)\right \Vert^2_{m,n}=\int_m |\lambda(x)|^2n(x)\mathrm{d}x $ with $n(x)$ being the probability density function which describes the region $m$. The combination of these terms gives the associated cost or loss function connected with the NN training. Thus, the objective function can be written as:

\begin{align}
\mathcal{L}(\zeta)&=\left\Vert \partial_t + \mathcal{A}-r)\zeta(t,S:\bm{\theta}) \right\Vert^2_{Y_1}+ \left\Vert \zeta(T,S:\bm{\theta})-g_1(T,S) \right\Vert^2_{Y_2}+ \left\Vert \zeta(t,0:\bm{\theta})-V_0(t) \right\Vert^2_{Y_3} \nonumber\\
& + \left\Vert \zeta(t,S:\bm{\theta})-g_2(t,S) \right\Vert^2_{Y_4}\,.
\end{align}

Suppose $\mathcal{L}(\zeta)=0$, then $\zeta(t,S:\bm{\theta})$  is a solution of the PDE in equation (\ref{c}). The major aim is to obtain a set of parameters $\bm{\theta}$ such that the function $\zeta(t,S:\bm{\theta})$ minimizes the observed error of $\mathcal{L}(\zeta)$. The procedure for seeking a good parameter set by minimizing this loss function using the stochastic gradient descent (SGD) optimizer is called ``training". Thus, using a Machine Learning approach, and in this case, the artificial NN, it will be feasible to minimize the function  $\mathcal{L}(\zeta)$ by applying the SGD approach on the sequence of asset and time points which are drawn at random. A part of the input samples, fully dependent on the mini-batch size, is drawn within each iteration to estimate the gradient of the given objective function. The gradient is then estimated over these mini-batches due to the limitations of the computer memory \cite{liu2019neural}.

The training of NN is classified into first identifying the network's hyperparameters, that is, the architectural structure and the loss function of the network. Next, use the SGD to optimize or estimate the loss minimizer, and then the derivatives of the loss function are computed using the backpropagation techniques. Essentially, the process of searching and identifying the best $\bm{\theta}$ parameter by minimizing the loss function using the gradient descent-based optimizers is referred to as ``training". The whole procedure is well illustrated in the algorithm below \cite{sirignano2018dgm}, and can be implemented in solving the PDE in equation (\ref{c}):

\begin{scriptsize}
\begin{algorithm}[H]
\caption{Implementing NN in solving PDE}
%\begin{algorithmic} 
\begin{enumerate}[(1)]
\item Set up the initial parameter set $\bm{\theta}_0=(\bm{W}_0,\bm{b}_0)$ as well as the learning rate $\alpha_n$, where $\bm{W}_0$ and $\bm{b}_0$ represent the weight and the bias respectively. 
\item Generate samples drawn randomly from the interior of the domain and from the time/asset boundaries. That is
\begin{itemize}
\item Generate $(t_n,S_n)$ from $[0,T] \times \Omega$ in accordance to the probability density $v_1$.
\item Generate $\omega_n$ from $\Omega$ in accordance to the probability density $v_2$.
\item Generate $r_n$ from $[0,T]$ in accordance to the probability density $v_3$.
\item Generate $(\tau_n,z_n)$ from $[0,T] \times \partial \Omega$ in accordance to the probability density $v_4$.
\end{itemize}
\item Estimate the value of squared error $\mathcal{G}(\theta_n; x_n)$ where $x_n$ is a set of randomly sampled points denoted by $x_n=\{(t_n,S_n), \omega_n, r_n, (\tau_n,z_n)\}$. That is, compute the following:

\begin{itemize}
\item Compute $\mathcal{G}_1(\theta_n; t_n, S_n) = \left( (\partial_t + \mathcal{A}-r)\zeta(\theta_n; t_n, S_n) \right)^2$.
\item Compute $\mathcal{G}_2(\theta_n; \omega_n) = \left( \zeta(T,\omega_n)-g_1(T,\omega_n)\right)^2$.
\item Compute $\mathcal{G}_3(\theta_n; r_n) =\left( \zeta(r_n,0)-V_0(r_n)\right)^2$.
\item Compute $\mathcal{G}_4(\theta_n; \tau_n, z_n) = \left( \zeta(\tau,z_n)-g_2(\tau_n,z_n)\right)^2$. 
\item Set $\mathcal{G}(\theta_n; x_n)= \mathcal{G}_1(\theta_n; t_n, S_n)+ \mathcal{G}_2(\theta_n; \omega_n)+ \mathcal{G}_3(\theta_n; r_n)+ \mathcal{G}_4(\theta_n; \tau_n, z_n)$.
\end{itemize}

\item Compute the gradient $\nabla_{\theta} \mathcal{G}(\theta_n; x_n)$ using backpropagation techniques.

\item Take an SGD step at the random points $x_n$ using the learning rates from the Adam optimization method. Use the estimated gradient to update the parameter  $\bm{\theta}_n$. Therefore, in each iteration $n$, the parameters $\bm{\theta}_n$ are updated in accordance to the relationship below:

\begin{equation}\label{d}
\theta_{n+1}=\theta_n - \alpha_n \nabla_{\theta} \mathcal{G}(\theta_n; x_n)
\end{equation}
Equation (\ref{d}) can be explicitly written in terms of the weights and bias as
\begin{equation}
\begin{cases}
               \bm{W}_{i+1}\leftarrow \bm{W}_{i} - \alpha_i \frac{\partial \mathcal{G}(\theta_i; x_i)}{\partial \bm{W} }\\
               \\
               \bm{b}_{i+1}\leftarrow \bm{b}_{i} - \alpha_i \frac{\partial \mathcal{G}(\theta_i; x_i)}{\partial \bm{b} }
            \end{cases}\,,
\end{equation}
where $i=0,1, \cdots, n$; and $n$ is the number of training iterations.

\item Repeat the procedure from (2-4) till the convergence property is satisfied, that is till $\left\Vert \theta_{n+1}-\theta_n  \right\Vert $ becomes very small.
\end{enumerate}
%\end{algorithmic}
\end{algorithm}
\end{scriptsize}

It must also be noted that the gradient descent steps $ \nabla_{\theta} \mathcal{G}(\theta_n; x_n)$ are the unbiased estimates of $ \nabla_{\theta} \mathcal{L}(\zeta(.; \theta_n))$, such that 
$$\mathbb{E}[ \nabla_{\theta} \mathcal{G}(\theta_n; x_n)| \theta_n] =  \nabla_{\theta}  \mathcal{L}(\zeta(.; \theta_n)),$$ and also
a negative correlation exists between the learning rate $\alpha_n$ and $n$.  \\

The learning rate $\alpha$ is employed to scale the intensity of the parameter updates during the gradient descent. Choosing the learning rate, as the descent parameter $\alpha$, in the network training is crucial since this factor plays a significant role in aiding the algorithm's convergence to the true solution in the gradient descent. It equally affects the pace at which the algorithm learns and whether or not the cost function is minimized. When $\alpha$ implies divergence, and thus, the optimal point can be missed. Also, a fixed value of $\alpha$ can most likely make the local optima overshoot after some iterations, leading to divergence. Defining the learning rate as a dynamic decreasing function is preferable since it allows the algorithm to identify the needed point rapidly. Another limitation in employing the gradient descent method is obtaining the value of the initial parameters of the loss function. If the value is significantly close to the local optimum, then the slope of the loss function reduces, thereby leading to the optimal convergence. Otherwise, there will be no convergence as the solution explodes abnormally.

\subsection{Solution of bitcoin option pricing PDE}

While the pricing PDE has a proven smooth solution under the model dynamics, obtaining the analytical form is intractable. Numerical methods must be used to approximate the solution. Neural networks provide a flexible parametric approach based on their universal approximation theoretical results. Given sufficient network capacity, neural networks can represent a wide class of functions, including solutions to PDEs like the pricing equation. However, challenges remain in finding the optimal network parameters to recover the true solution robustly. While existence is guaranteed, the non-convex optimization of neural networks does not assure convergence to global optimality. Care must be taken in specifying the network architecture, loss function, regularization, and training methodology to promote the learning of the pricing function. Subject to these caveats, neural networks present a promising computational technique for approximating smooth pricing solutions, circumventing discretization of the domain. We note the limitations of using neural networks as function approximators. However, their generalization capabilities provide a methodology for data-driven extraction of the pricing mapping across the entire domain. This avoids the constraints of methods like grid-based techniques that rely on local consistency. Future work should explore neural network training enhancements and theoretical guarantees to ensure robust solutions.\\

Thus, from equation (\ref{b}), let $V$ be the price of the bitcoin call option with strike price $E$ and $S'$ be the price of the bitcoin option at time $t'$. Let $r$ be the risk-free rate; $\sigma_*$, the volatility of the bitcoin; $T$, the expiration of the contract and denote bitcoin call price $V=V(t',S')$ to be a function of the remaining time to maturity and the bitcoin price. Considering the European call option, the terminal or the final condition and the boundary conditions are denoted respectively by:
\begin{equation}\label{e}
\begin{cases}
               TC: V(T,S')=\max\{S'-E,0\} \\
               \\
               BC: V(t',0)=0\\
               BC: \frac{V(t',S')}{S'} \rightarrow 1 \qquad \text{for} \quad S' \rightarrow \infty \,,
            \end{cases}
\end{equation}  
 
for $t' \in [0,T]$ and $S' \in [0, S_{\max}]$, where $S_{\max}$ is the upper limit\footnote{We note that since crypto markets are not regulated, the prices may deviate very largely and the maximum could be infinity. However, since we are considering option pricing with European features, we chose the maximum price that the bitcoin has attained from our dataset.} which the bitcoin can accumulate prior to the option's expiration. With the change of variables $t=\frac{T-t'}{T}$, and $S=\frac{S'}{S_{\max}}$, such that $t \in [1,0]$ and $S \in [0,1]$, we can substitute $- T \partial t = \partial t'$ into the PDE (\ref{b}). This restriction is imposed such that the terminal condition (TC) in equation (\ref{e}) can be transformed into an initial condition (IC). Thus, the PDE reduces to 

\begin{equation}\label{f}
\frac{1}{T}\frac{\partial V}{\partial t} - \frac{\sigma_*^2 S^2}{2} \frac{\partial^2 V}{\partial S^2} - \eta S\frac{\partial V}{\partial S} +rV + \beta(S)=0\,,
\end{equation}
with the following transformed conditions:

\begin{equation}\label{g}
\begin{cases}
               IC: V(0,S)=\max\left\{S- \frac{E}{S_{\max}},0\right\}, \quad \forall S \in [0,1] \\
               \\
               BC: V(t,0)=0, \quad \forall t \in [1,0]\\
               BC: V(t,1)=1- \frac{E}{S_{\max}} \,.
            \end{cases}
\end{equation}   

Solving the PDE in equation (\ref{f}) using the conditions in equation (\ref{g}) entails introducing a trial solution $\zeta(t,S:\bm{\theta})$ which employs the feedforward NN with parameter $\bm{\theta}$ corresponding to the weight and biases of the network's architecture. The trial function $\zeta(t,S:\bm{\theta})$ of the NN is constructed such that it satisfies the boundary and the initial conditions \cite{eskiizmirliler2020solution, lagaris1998artificial, yadav2015introduction}. This can be written as 

\begin{equation}\label{h}
\zeta(t,S:\bm{\theta}) = A(t, S) + B(t, S)N(t, S: \bm{\theta})\,,
\end{equation} 

where $B(t,S)=tS (1-S)$. The first term $A(t, S)$ satisfies the boundary and the initial conditions has no adjustable parameters and is denoted by 
\[A(t,S)= (1-t) \max \left\{S-\frac{E}{S_{\max}},0\right\} + tS\left(1-\frac{E}{S_{\max}}\right)\,. \]
The last term in equation (\ref{h}) with the NN function has adjustable parameters, and it handles the minimization problem. This term does not contribute to the boundary condition because at boundaries $S=0, S=1$ and $t=0$; the term became exactly zero. Thus, the trial function is explicitly given as

\begin{equation}\label{i}
\zeta(t,S:\bm{\theta}) = (1-t) \max \left \{S-\frac{E}{S_{\max}},0\right\} + tS\left(1-\frac{E}{S_{\max}}\right) + tS (1-S)N(t, S: \bm{\theta})\,.
\end{equation} 

Given the input values $(t, S)$, the network's output is the net function which is defined by 

\begin{equation}\label{j}
N(t, S: \bm{\theta}) = \sum_{r=1}^M q_r f(w_r t + \gamma_r S + b_r)\,,
\end{equation}

where $M$ is the total number of the neurons which are present in the hidden layer of the NN; $f(z_r)$ is the activation function; $q_r$ is the synaptic weight of the $r^{\text{th}}$ hidden neuron to the output; $b_r$ is the bias value of the $r^{\text{th}}$ hidden neuron; $w_r$ and $\gamma_r$ are the synaptic coefficients from the time input to the $r^{\text{th}}$ hidden neuron and from the spatial inputs to the $r^{\text{th}}$ hidden neuron respectively.\\

Finally, with regards to the cost function, we first discretise the domains to convert the PDE in equation (\ref{f}) which are subject to the boundary conditions, into an unconstrained minimization problem. For $S \in [0, 1]$ and $t \in [1,0]$, consider $\Delta S$ to be a uniform step size for the bitcoin asset $S$, and $\Delta t$ to be the step size for the time $t$, such that
\begin{align*}
S_i&=i \Delta S \implies \Delta S = \frac{1}{N_s}, \qquad \text{for} \quad i=0,1,\cdots, N_s \qquad \text{and} \\
t_j&=j \Delta t \implies \Delta t = \frac{1}{N_t}, \qquad \text{for} \quad j=0,1,\cdots, N_t \,,
\end{align*}

where $N_s$ and $N_t$ are the step sizes for the bitcoin asset and the time, respectively. The dataset that will be generated from this discretised domain will consist of a matrix of size $(N_s \times N_t) \times 2$, that is, two columns made up by $S_i$ and $t_j$, as well as $(N_s \times N_t) $ number of rows. The dataset will be split into the train data and the test dataset and the NN will be trained on the train dataset and tested on a separate dataset.\\

Using the trial function in equation (\ref{i}), denote the bitcoin option value at time $t_j$ for the asset $S_i$ given by $\zeta(t_j,S_i:\bm{\theta})=\zeta(j\Delta t,i \Delta S:\bm{\theta})$ as $\zeta_{j,i}$, then the cost function which the ANN has to minimize is

\begin{equation}\label{k}
L(\bm{\theta})= \frac{1}{2}\sum_{j=1}^{N_t}\sum_{i=1}^{N_s} \left[ \frac{\partial \zeta_{j,i}}{\partial t_j} - \frac{\sigma_*^2 S_i^2}{2} \frac{\partial^2 \zeta_{j,i}}{\partial S_i^2} - \eta S_i\frac{\partial \zeta_{j,i}}{\partial S_i} +r\zeta_{j,i} + \beta(S_i) \right]^2\,, 
\end{equation}
for $j =0,1,\cdots, N_t$ and $i=0,1,\cdots, N_s$.\\

Essentially, the loss function is minimized during the training phase to determine the optimal NN parameters. The partial derivatives in equation (\ref{k}) are computed as follows:

\begin{align}
\frac{\partial \zeta}{\partial t}&= - \max \left\{S-\frac{E}{S_{\max}},0 \right\} + S\left(1-\frac{E}{S_{\max}} \right) + tS(1-S)\frac{\partial N(t,S: \bm{\theta})}{\partial t} + S(1-S)N(t,S: \bm{\theta})\label{v1} \\&\nonumber\\
\frac{\partial \zeta}{\partial S}&= t\left(1-\frac{E}{S_{\max}} \right) + tS(1-S)\frac{\partial N(t,S: \bm{\theta})}{\partial S} + N(t,S: \bm{\theta})(t-2tS),  \quad \text{for} \; S \leq \frac{E}{S_{\max}}\label{v2} \\&\nonumber\\
\frac{\partial \zeta}{\partial S}&= (1-t) + t\left(1-\frac{E}{S_{\max}} \right) + tS(1-S)\frac{\partial N(t,S: \bm{\theta})}{\partial S} + N(t,S: \bm{\theta})(t-2tS), \quad \text{for} \; S > \frac{E}{S_{\max}}\label{v3}    \\&\nonumber\\
\frac{\partial^2 \zeta}{\partial S^2}&= tS(1-S)\frac{\partial^2 N(t,S: \bm{\theta})}{\partial S^2} + 2t(1-2S) \frac{\partial N(t,S: \bm{\theta})}{\partial S} +-2tN(t,S: \bm{\theta})\,.  \label{v4}       
\end{align}
 
The partial derivatives of the NN with respect to $t,S$ and $S^2$ from equations (\ref{v1}-\ref{v4}) are dependent on the nature of the activation function used. For instance, if we consider the sigmoid activation function, then $f(z_r)=(1+ \mathrm{e}^{-z_r})^{-1}$. Then from equation (\ref{j}), $z_r= w_r t + \gamma_r S + b_r$, and thus, the NN function in expanded form can be written as
\begin{equation}\label{js}
N(t, S: \bm{\theta}) = \sum_{r=1}^M q_r\left(1+ \mathrm{e}^{-(w_r t + \gamma_r S + b_r)}\right)^{-1} \,,
\end{equation}

\noindent Taking the derivatives of equation (\ref{js}) with respect to $t,S$ and $S^2$, we have the following
\begin{align}
\frac{\partial N(t,S: \bm{\theta})}{\partial t}&=\sum_{r=1}^M q_r w_r f(z_r)(1-f(z_r)) \\&\nonumber\\
\frac{\partial N(t,S: \bm{\theta})}{\partial S}&=\sum_{r=1}^M q_r \gamma_r f(z_r)(1-f(z_r))\\&\nonumber\\
\frac{\partial^2 N(t,S: \bm{\theta})}{\partial S^2}&=\sum_{r=1}^M q_r \gamma_r^2 f(z_r)(1-f(z_r))(1-2f(z_r)) \,.         
\end{align}
where $z_r= w_r t + \gamma_r S + b_r$. For the tanh activation function, $f(z_r)=(1-\mathrm{e}^{-2z_r})(1+\mathrm{e}^{-2z_r})^{-1}$, and the derivatives of the NN are also easily obtained mathematically. However, when the ReLU\footnote{The tensorflow sigmoid and the ReLU function were used in this research.} activation $f(z_r)=\max\{0,z_r\}$ is used, the mathematical NN derivatives become complicated due to the Heaviside step function and the presence of dirac delta in the first and second derivatives, respectively. In the context of backpropagation of error, these derivatives are needed when the weights of the NN are constantly updated. The only assumption is the value zero which is obtained when the derivative at point zero is taken.
 
\section{Numerical results, implementation and discussion}
This section introduces the empirical and analytical structure, approximation of parameters, the estimation of option prices, as well as the limitations of the efficiency and no-arbitrage assumptions.

\subsection{Empirical Analysis}
\subsubsection{Data Source}
For the cryptocurrency data source, we used the bitcoin historical closing prices on the CoinGecko website \footnote{CoinGecko provides a fundamental analysis of the structure of the digital currency market; The data was accessed on August 16, 2021, at \url{https://www.coingecko.com/en}} based on US dollars, covering five years from August 1, 2016, till July 31, 2021. As of access, the global crypto market capitalization of \$2.09 Trillion and the CoinGecko tracks 8,934 cryptocurrencies, with 42.2\% dominance for bitcoin. Furthermore, we used the Google trend data \footnote{\url{https://trends.google.com/trends/explore?q=bitcoin}} to extract the bitcoin sentiment data. The Google trend data provides a scaled time series of the number of times bitcoin has been searched, so the maximum is 100. Figures \ref{Bitc} and \ref{sent} describe the dataset for the bitcoin prices and the corresponding sentiment trends, respectively. 

\subsubsection{Descriptive Statistics}

\begin{figure}[H]
  \begin{subfigure}[b]{0.5\textwidth}
    \includegraphics[width=\textwidth]{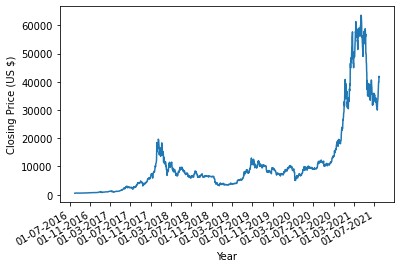}
    \caption{\footnotesize{Bitcoin closing prices}}
    \label{Closing}
  \end{subfigure}
  \hfill
  \begin{subfigure}[b]{0.5\textwidth}
    \includegraphics[width=\textwidth]{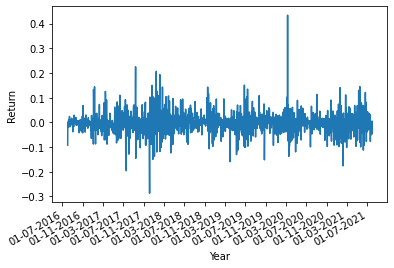}
    \caption{\footnotesize{Bitcoin returns}}
    \label{Returns}
  \end{subfigure}
  \caption{Bitcoin daily closing prices and log returns during 2016-2021}
  \label{Bitc}
\end{figure}

The dataset is sampled on a daily frequency, and the results are plotted in Figure \ref{Bitc}. We used the adjusted bitcoin closing prices to estimate the continuously compounded returns. For the log-return $r_i$ at time $t_i$, we used the expression $r_i = \log \left(\frac{S_{i}}{S_{i-1}}\right)$, where $S_i$ is the bitcoin price at time $t_i$. Since bitcoin is traded daily, we use the daily sample data, giving 365 observations per year, whereas the trend data is sampled weekly. The descriptive statistics of the dataset used in this work are presented in Table \ref{DS}.  

\begin{figure}[H]
  \begin{subfigure}[b]{0.5\textwidth}
    \includegraphics[width=\textwidth]{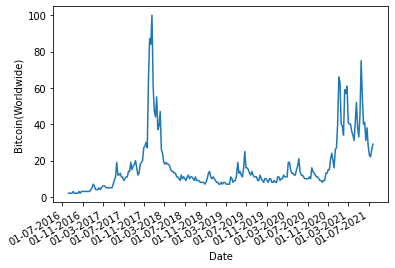}
    \caption{\footnotesize{Values}}
    \label{Closing1}
  \end{subfigure}
  \hfill
  \begin{subfigure}[b]{0.5\textwidth}
    \includegraphics[width=\textwidth]{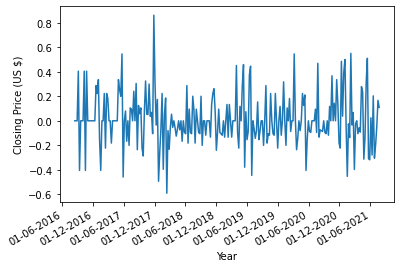}
    \caption{\footnotesize{Returns}}
    \label{Returns1}
  \end{subfigure}
  \caption{Sentiment data for bitcoin during 2016-2021}
  \label{sent}
\end{figure}

Both bitcoin prices and the corresponding log-returns react to big events in the cryptocurrency market. From Figure \ref{Closing}, a considerable surge in bitcoin prices was observed after March 2017, owing to the widespread interest in cryptocurrencies. This jump was later affected by a series of political interventions leading to a drop in June 2017 \cite{hou2020pricing}. Furthermore, in late 2020 and early 2021, another dramatic increase in the prices was seen due to the increased acquisition by large investors, financial institutions and corporations. These intensive movements in the cryptocurrency markets have been captured extensively by the corresponding bitcoin sentiment data as found in Figure \ref{Closing1}.

%\begin{table}[H]
%\center
%\caption{Descriptive statistics}
%\label{DS}
%\begin{tabular}{|p{4cm}cc|}\hline
% \hline 
% \textbf{Statistic} & \textbf{Bitcoin closing prices} & \textbf{Sentiment values}  \\ 
% \hline \hline
%No of observations & 1826 & 260\\ 
%Mean & 11218.89112 & 16.82692\\  
%Minimum & 517.13000 & 2.00000 \\
%Q1 & 3820.72500 & 8.75000 \\
%Median & 7534.74000 & 11.00000 \\
%Q3 & 10670.02750 & 18.25000 \\
%Maximum & 63577.00000 & 100.00000 \\   
%Standard deviation & 13177.87426 & 15.70766\\ 
%Skewness & 2.25804 & 2.37818\\ 
%Kurtosis & 4.49190 & 6.50161 \\ 
% \hline \hline
% \end{tabular}  
%\end{table}

\begin{table}[H]
\center
\caption{Descriptive statistics for the log-returns}
\label{DS}
\begin{tabular}{|p{4cm}cc|}\hline
 \hline 
 \textbf{Statistic} & \textbf{Bitcoin closing prices} & \textbf{Sentiment values}  \\ 
 \hline \hline
No of observations & 1825 & 259\\ 
Mean & -0.00241 & 0.01033\\  
Minimum & -0.28701 & -0.59205 \\
Q1 & -0.01969 & -0.10536 \\
Median & -0.00251 & 0.00000 \\
Q3 & 0.01336 & 0.10728 \\
Maximum & 0.43371 & 0.86305 \\   
Standard deviation & 0.04132 & 0.20934\\ 
Skewness & 0.67119 & 0.45648\\ 
Kurtosis & 10.53706 & 1.20553 \\ 
 \hline \hline
 \end{tabular}  
\end{table}

Table \ref{DS} presents the descriptive statistics for the log-returns on the bitcoin closing prices, as well as the sentiment data values. The dual-dataset consists of a sample of 260 weekly sentiment values and 1826 bitcoin daily closing prices. The Skewness/Kurtosis test is one of three common normality tests designed to detect all the deviations from normality and determine the shape of the distribution for the dataset. For a normal distribution, the skewness is zero, and the kurtosis is three. From the table, the log-return for the dataset of the sentiment and the bitcoin closing prices are positively and highly skewed to the right. With the kurtosis of 1.20553 and 10.53706, the two datasets have heavier, longer and fatter tails than the normal distribution, and they can be referred to as leptokurtic. 

\subsubsection{Parameter Estimation}\label{pa}
In this section, we estimate the parameters for the numerical computation, and the results finally presented are the values of the bitcoin options, having the European call features. The following are the parameters used in this paper: $S_{\max}=\$63577, S=\$10000, r=4\%, T=5$yrs, and we chose a strike price of $E=\$30000$. We used the mean and variance of the log-return from the sentiment index data as $\mu_p=0.01033$ and $\sigma_p=0.20934$, respectively. Also, from the log-return of the bitcoin closing prices, we calculated the mean and the variance of the log-return as $\mu_d=-0.00241$ and $\sigma_d=0.04132$, respectively. Next, we estimate the $\lambda$, the jump intensity rate, together with $k$, the expectation of the relative price of the jump size, since it is essential to decide when a jump occurs. This parameter estimation was done using the maximum likelihood estimation method since there are no closed-form expressions for the optimal values of these parameters. Also, the daily bitcoin price return is measured in years, that is, $\Delta t \sim dt=1/365=0.00274$. \\

Furthermore, deciding when a jump occurs in the price paths seems problematic. We adopt the techniques of \citep{tang2018merton} and Hanson \& Zhu (2004)\citep{hanson2004comparison} to estimate the parameters of the jump-diffusion models, who suggested a specific threshold $\epsilon$ with the aim of determining whether a jump has occurred or not. In this case, maximum likelihood estimation is not strongly dependent on the value of the threshold $\epsilon$ \citep{tang2018merton}. Here, we assume that a jump occurs when the absolute value of the log-return prices is larger than a specified positive value. The intensity rate $\lambda$ is measured as 
\[\lambda = \frac{\text{number of jumps}}{\text{period length in years}}= \text{number of jumps per year}\]
For our estimate, we set the threshold level $\epsilon=0.07$. If $\epsilon$ is too small, then the majority of the price movements would be considered jumps. On the other hand, if $\epsilon$ is large, then the set of absolute jump size $y_t$ could be empty, thereby making the parameters to be estimated would be undetermined. Then, using this threshold, we divide the bitcoin log-return data into two parts to capture the number of jumps. The first part captures the values when the absolute value of the log-return is greater than $\epsilon$, and we assume that a jump occurred here. The other part consists of no jump, and it captures the values whose log-return is lesser than $\epsilon$. From the techniques, we obtained the remaining parameters $\lambda=31.8$ and $k=\mathbb{E}[J_t]=-0.002195$.

\subsection{Numerical Implementation}
For the NN architecture, we employed the random search method to obtain the optimal hyperparameter. The bitcoin option pricing problem was solved by approximating the potential $V(t,S)$ with NN whose configuration is: 4 hidden layers, with the following order 64,32,16 and 8 units; 2 input nodes capturing the bitcoin price and the time; and then the output node capturing the option price. (Hence, the configuration 2-64-32-16-8-1). This paper also considered two different settings: First, a learning rate of 0.001, iteration step of 10000, a sigmoid activation function\footnote{Sigmoid is defined by $f(z)=1/(1+\exp{(-z)})$, where $z$ is the input to the neuron} and the SGD -  stochastic gradient descent optimizer (Model I). Secondly, a learning rate of 0.001, iteration step of 10000, a ReLU activation function\footnote{ReLU is defined by $f(z)=\max [0,z]$, where $z$ is the input to the neuron} and the Adam optimizer\footnote{Adam - Adaptive Moment Estimation is a stochastic optimization method which is based on adaptive estimates of lower-order moments. They can be applied to solving non-convex optimization problems in machine learning (See \url{https://arxiv.org/pdf/1412.6980v9.pdf}).} (Model II). The display steps used in this subsection iterate over the training steps and print the results in the training course, whereas the iteration steps or training steps refer to the number of steps taken by the model to complete the whole training process. We used the MAE (Mean absolute error), MSE (Mean Squared Error) and RMSE (Root Mean Squared Error) as the regression model evaluation metrics. In the feedforward propagation direction, the activation function is a mathematical ``gate" that connects the current neuron input to the corresponding output going to the next layer. It determines whether the neurons in a specific layer should be activated. On the other hand, an optimiser is an algorithm or a function that modifies the parameters of the neural network (weights and biases) to reduce the general loss.

\subsubsection{Model I}
For comparative purposes, we considered the impact of using the SGD optimizer with the sigmoid activation function on the loss and option values for both the Black-Scholes model and the jump Merton diffusion models. The tables below give the standard evaluation metrics in terms of the MSE and RMSE (Table \ref{loss}), as well as the MAE (Table \ref{lossmae}) for the proposed models.

\begin{table}[H]
\center
\caption{Model I - Loss values (MSE; RMSE) and iteration numbers}
\label{loss}
\begin{tabular}{|cccccc|}\hline
 \hline 
\textbf{Display Steps} & \multicolumn{5}{c|}{\textbf{Loss Values}} \\
&&&&&\\
   & \multicolumn{2}{c}{\textbf{Black-Scholes Model}} && \multicolumn{2}{c|}{\textbf{Jump Merton Diffusion Model}} \\ 
\hline\hline
   &$(10 \times 10)$ grid & $(20\times 20)$ grid   && $(10 \times 10)$ grid & $(20\times 20)$ grid  \\ 
    &&&&& \\ 
   &MSE; RMSE & MSE; RMSE   && MSE; RMSE & MSE; RMSE  \\ 
 \hline \hline
 
1  & 0.60858; 0.78012    &0.14921; 0.38628  && 110.39897; 10.50710  & 63.95981; 7.99749 \\
500 & 0.42021; 0.64824  &0.10601; 0.32559  && 86.10558; 9.27930  & 44.13830; 6.64367\\
1000 & 0.29216; 0.54052 &0.07549; 0.27476 && 62.54251; 7.90838 & 30.97948; 5.56592\\
1500 & 0.20463; 0.45236 &0.05390; 0.23216 && 46.75750; 6.83795 & 23.03805; 4.79980\\
2000 & 0.14428; 0.37984 &0.03859; 0.19644 && 37.76791; 6.14556 & 18.51702; 4.30314\\
2500 & 0.10234; 0.31991 &0.02770; 0.16643 && 32.73466; 5.72142& 16.00801; 4.00100\\
3000 & 0.07297; 0.27013 &0.01995; 0.14125 && 29.91682; 5.46963& 14.63013; 3.82494\\
3500 & 0.05229; 0.22867 &0.01442; 0.12008 && 28.33798; 5.32334& 13.87676; 3.72515\\
4000 & 0.03766; 0.19406 &0.01048; 0.10237 && 27.45284; 5.23955& 12.46562; 3.53067\\
4500 & 0.02727; 0.16514 &0.00767; 0.08758 && 26.95636; 5.19195& 12.04141; 3.47007\\
5000 & 0.01988; 0.14100 &0.00566; 0.07523 && 26.67778; 5.16505& 11.11913; 3.33454\\
5500 & 0.01461; 0.12087 &0.00422; 0.06496 && 26.52140; 5.14989& 10.05239; 3.17055\\
6000 & 0.01084; 0.10412 &0.00320; 0.05657 && 25.43356; 5.04317& 9.61590; 3.10095\\
6500 & 0.00814; 0.09022 &0.00246; 0.04960 && 24.38417; 4.93803& 9.19589; 3.03247\\
7000 & 0.00620; 0.07874 &0.00194; 0.04405 && 22.35636; 4.72825& 8.98484; 2.99747\\
7500 & 0.00482; 0.06943 &0.00156; 0.03950 && 21.84065; 4.67340& 8.07869; 2.84230\\
8000 & 0.00382; 0.06181 &0.00129; 0.03592 && 20.33174; 4.50907& 7.77519; 2.78840\\
8500 & 0.00310; 0.05568 &0.00110; 0.03317 && 19.62665; 4.43020& 5.97315; 2.44400\\
9000 & 0.00259; 0.05089 &0.00096; 0.03098 && 18.32369; 4.28062& 5.37190; 2.31774\\
9500 & 0.00221; 0.04701 &0.00086; 0.02933 && 16.92194; 4.11363& 4.97108; 2.22959\\
10000 & 0.00195; 0.04416 &0.00079; 0.02811 && 16.32086; 4.03991& 4.16051; 2.03973\\
\hline
 \hline
 \end{tabular}  
\end{table}

\begin{table}[H]
\center
\caption{Model I - MAE Loss values and iteration numbers}
\label{lossmae}
\begin{tabular}{|cccccc|}\hline
 \hline 
\textbf{Display Steps} & \multicolumn{5}{c|}{\textbf{MAE Loss Values}} \\
&&&&&\\
   & \multicolumn{2}{c}{\textbf{Black-Scholes Model}} && \multicolumn{2}{c|}{\textbf{Jump Merton Diffusion Model}} \\ 
\hline\hline
   &$(10 \times 10)$ grid & $(20\times 20)$ grid   && $(10 \times 10)$ grid & $(20\times 20)$ grid  \\ 
 \hline \hline
 
1  & 0.72932   &0.04738  && 48.67608  & 23.95461 \\
500 & 0.03567  &0.02418  && 36.67128  & 18.45554 \\
1000 & 0.03342 &0.02206 && 28.45487 & 15.76733\\
1500 & 0.03177 &0.02091 && 22.93645 & 13.55667\\
2000 & 0.02533 &0.02002 && 19.32654 & 10.58556\\
2500 & 0.02045 &0.01871 && 17.03814 & 10.26077\\
3000 & 0.01918 &0.01613 && 16.64287 & 10.20143\\
3500 & 0.01773 &0.01011 && 15.81879 & 10.18432\\
4000 & 0.01646 &0.00636 && 15.67284 & 10.00146\\
4500 & 0.01493 &0.00571 && 15.07281& 9.99704\\
5000 & 0.01110 &0.00539 && 14.92092& 9.70993\\
5500 & 0.00688 &0.00504 && 14.83609& 9.68770\\
6000 & 0.00620 &0.00461 && 14.65356& 9.61490\\
6500 & 0.00607 &0.00423 && 14.60443& 9.59044\\
7000 & 0.00593 &0.00399 && 14.51234& 9.56771\\
7500 & 0.00589 &0.00375 && 14.44422& 8.99087\\
8000 & 0.00586 &0.00364 && 14.40773& 8.85401\\
8500 & 0.00583 &0.00358 && 14.19245& 8.69760\\
9000 & 0.00580 &0.00352 && 13.99869& 8.49980\\
9500 & 0.00575 &0.00348 && 13.97268& 8.47443\\
10000 & 0.00571 &0.00343 && 13.97001 & 8.45389\\
\hline
 \hline
 \end{tabular}  
\end{table}

Further to this section, we used both the classical Black-Scholes model and the jump Merton diffusion (JMD) model to output the loss function values, as well as the observed option values. For the Black-Scholes, we used the relevant PDE, by equating $\beta(S)=0$ and $\eta=r$ in equation (\ref{f}) subject to the conditions in equation (\ref{g}). During the training phase of the NN, we aim to reduce the error or the cost function in equation (\ref{k}) as small as possible to achieve an efficient optimization technique. Table \ref{loss} gives the loss values for the two models. In each model, we partitioned the asset (bitcoin closing prices) and the time into 10 and 20 uniform grid spaces to investigate the nature of the loss values. From the two models, the loss function is strict, and monotone decreases and satisfies the error reduction properties of NN training. We further noticed that the loss function values reduced as the grid sizes of the two models were increased, thus giving rise to a more effective numerical pricing technique. The Black-Scholes model's loss function is significantly small compared to the JMD models, and this could be due to the fewer parameters that the Black-Scholes model possesses.

\begin{figure}[H]
  \begin{subfigure}[b]{0.5\textwidth}
    \includegraphics[width=\textwidth]{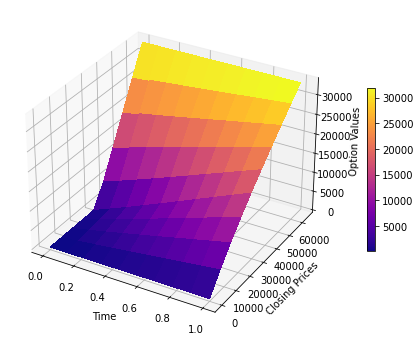}
    \caption{3D-View: Black-Scholes Prices}
    \label{bs3d}
  \end{subfigure}
  \hfill
  \begin{subfigure}[b]{0.5\textwidth}
    \includegraphics[width=\textwidth]{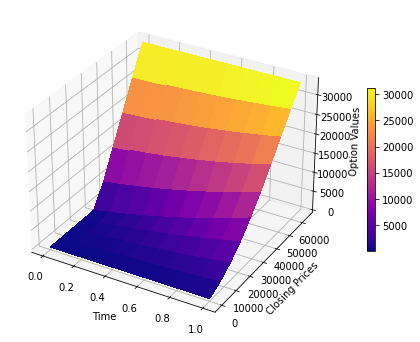}
    \caption{3D-View: Merton Jump Diffusion Prices}
    \label{mjd3d}
  \end{subfigure}
   \caption{3-dimensional option plots for Black-Scholes and Merton jump-diffusion model prices -- Model I}
    \label{3dplot}
\end{figure}

\begin{figure}[H]
  \begin{subfigure}[b]{0.5\textwidth}
    \includegraphics[width=\textwidth]{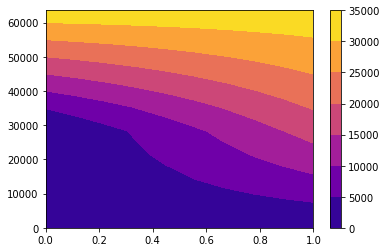}
    \caption{2D-View: Black-Scholes Prices}
    \label{bs2d}
  \end{subfigure}
  \hfill
  \begin{subfigure}[b]{0.5\textwidth}
    \includegraphics[width=\textwidth]{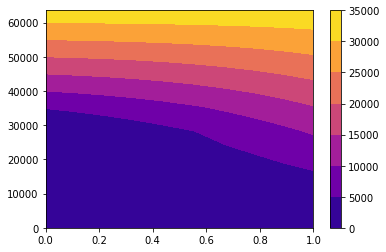}
    \caption{2D-View: Merton Jump Diffusion Prices}
    \label{mjd2d}
  \end{subfigure}
        \caption{2-dimensional option plots for Black-Scholes and Merton jump-diffusion model prices -- Model I}
        \label{2dplot}
\end{figure}

Figures \ref{3dplot} and \ref{2dplot} give 3-dimensional and 2-dimensional plots of the bitcoin call option values, respectively,  when the asset price process is modelled using both the Black-Scholes model and the MJD models. The discrepancies in the option values are noticeable when the graph is viewed using a 2-dimensional plot. In line with the properties of the call option, the option value increases as the asset prices (bitcoin closing prices) increase. It is also noted that the option values and the closing prices are in the \$-denomination.

\begin{figure}[H]
  \begin{subfigure}[b]{0.5\textwidth}
    \includegraphics[width=\textwidth]{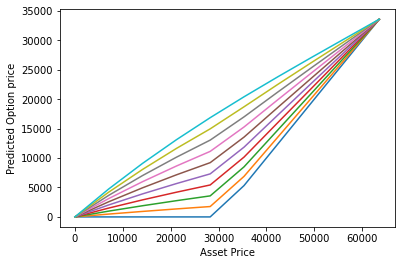}
    \caption{Black-Scholes Prices}
    \label{bso}
  \end{subfigure}
  \hfill
  \begin{subfigure}[b]{0.5\textwidth}
    \includegraphics[width=\textwidth]{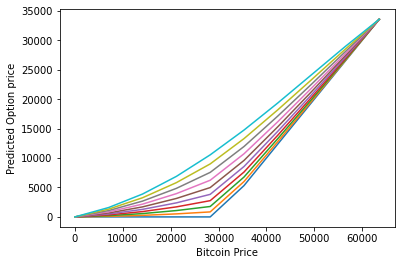}
    \caption{Merton Jump Diffusion Prices}
    \label{mjdo}
  \end{subfigure}
        \caption{Model I - Option price plots for Black-Scholes and Merton jump-diffusion models}
        \label{optionv}
\end{figure}

To have a clearer view of the nature of the option price concerning the closing prices, we plot the results obtained in Figure \ref{optionv}. The discontinuity at the strike price $E= \$30,000$ is observed, as the option remains out-of-the-money when the asset price is lesser than the strike price. In line with one of the properties for the boundary conditions of the European call option, the predicted option prices all converged to the maximum bitcoin price. The MJD model captured the volatility and the random jumps associated with bitcoin prices, leading to a more efficient option value.

\begin{table}[H]
\center
\caption{Model I -- Option values using the $10 \times 10$ grid for different uniform time-grid}
\label{optionval}
\begin{tabular}{|cccccccc|}\hline
 \hline 
 \textbf{Bitcoin Prices (\$)} & \multicolumn{7}{c|}{\textbf{Option Values (\$)}}\\
   & \multicolumn{3}{c}{\textbf{Black Scholes Model}} && \multicolumn{3}{c|}{\textbf{Jump Merton Diffusion}} \\ 
&&&&&&&\\
   &$t_1=0.333$ & $t_2=0.667$ & $t_3=1.000$   && $t_1=0.333$ & $t_2=0.667$ & $t_3=1.000$  \\ 
 \hline \hline
 
7064  & 1504.31  &3121.24  &4748.86  && 347.89  & 867.38   & 1600.07 \\
14128 & 2894.17  &5985.07  &9093.49  && 904.94  & 2169.36  & 3887.00\\
21192 & 4194.14  &8640.36  &13106.96 && 1703.56 & 3953.88  & 6883.78\\
28256 & 5426.56  &11131.75 &16856.50 && 2761.60 & 6227.06  & 10543.03\\
35320 & 10158.48 &15272.99 &20403.29 && 7624.26 & 10724.48 & 14754.47\\
42384 & 16022.81 &19907.55 &23802.45 && 13884.06& 16175.83 & 19359.16\\
49449 & 21872.43 &24284.63 &27103.13 && 20336.41& 21890.31 & 24166.82\\
56513 & 27720.24 &29034.40 &30348.71 && 26923.51& 27737.43 & 28973.52\\
63577 & 33577.00 &33577.00 &33577.00 && 33577.00& 33577.00 & 33577.00\\
\hline
 \hline
 \end{tabular}  
\end{table}

Table \ref{optionval} explicitly gives the option values for the two models, as it considers the $(10 \times 10)$ mesh sizes for both the asset and time parameters. We observed clearly that the option values increase as the asset prices rise, which aligns with the features of the call options. Furthermore, using the randomly selected time-grid of $t_1=0.333, t_2=0.667, t_3=1.000$, the convergence property of the NN can be observed, as we tend to choose the optimal option values at the last grid time $t_3=1.000$. 

\subsubsection{Model II}
This model uses a slightly different network configuration and the difference from Model I is the presence of the ReLU activation function and the Adam optimizer. We further obtain the standard evaluation metrics in terms of the MSE and RMSE (Table \ref{losss}), as well as the MAE (Table \ref{lossmaee}) for the proposed models.

\begin{table}[H]
\center
\caption{Model II - Loss values (MSE; RMSE) and iteration numbers}
\label{losss}
\begin{tabular}{|cccccc|}\hline
 \hline 
\textbf{Display Steps} & \multicolumn{5}{c|}{\textbf{Loss Values}} \\
&&&&&\\
   & \multicolumn{2}{c}{\textbf{Black-Scholes Model}} && \multicolumn{2}{c|}{\textbf{Jump Merton Diffusion Model}} \\ 
\hline\hline
   &$(10 \times 10)$ grid & $(20\times 20)$ grid   && $(10 \times 10)$ grid & $(20\times 20)$ grid  \\ 
    &&&&& \\ 
   &MSE; RMSE & MSE; RMSE   && MSE; RMSE & MSE; RMSE  \\ 
 \hline \hline
 
1  & 6.74934; 2.59795   &0.10188; 0.31919  && 92.69813; 9.62799   & 229.85840; 15.16108  \\
500 & 0.03997; 0.19993  &0.00182; 0.04266  && 26.29302; 5.12767   & 62.42821; 7.90115 \\
1000 & 0.01091; 0.10445 &0.00170; 0.04123 && 23.43724; 4.84120  & 57.07428; 7.55267 \\
1500 & 0.00680; 0.08246 &0.00124; 0.03521 && 21.61147; 4.64882  & 55.49834; 7.44972 \\
2000 & 0.00222; 0.04712 &0.00100; 0.03162 && 20.28269; 4.50363  & 53.79785; 7.33470 \\
2500 & 0.00113; 0.03362 &0.00087; 0.02950 && 19.43050; 4.40800  & 51.41717; 7.17058 \\
3000 & 0.00099; 0.03146 &0.00050; 0.02236 && 18.91586; 4.34923  & 48.40485; 6.95736 \\
3500 & 0.00079; 0.02811 &0.00032; 0.01789 && 18.55398; 4.30743 & 44.95583; 6.70491 \\
4000 & 0.00059; 0.02429 &0.00018; 0.01342 && 18.28303; 4.27587 & 41.20977; 6.41948\\
4500 & 0.00042; 0.02049 &0.00011; 0.01049 && 18.08552; 4.25271 & 37.84912; 6.15216\\
5000 & 0.00033; 0.01817 &0.00009; 0.00949 && 17.80039; 4.21905 & 34.43061; 5.86776\\
5500 & 0.00026; 0.01613 &0.00005; 0.00707 && 17.33213; 4.16319 & 32.10260; 5.66592\\
6000 & 0.00024; 0.01549 &0.00003; 0.00548 && 16.54747; 4.06786 & 30.05283; 5.48205\\
6500 & 0.00017; 0.01304 &0.00003; 0.00548 && 15.29592; 3.91100 & 28.33120; 5.32271\\
7000 & 0.00014; 0.01183 &0.00002; 0.00447 && 14.04497; 3.74766 & 26.53446; 5.15116\\
7500 & 0.00009; 0.00949 &0.00002; 0.00447 && 13.41838; 3.66311 & 24.63345; 5.16076\\
8000 & 0.00007; 0.00837 &0.00002; 0.00447 && 13.29875; 3.64675 & 23.66058; 4.86422\\
8500 & 0.00006; 0.00775 &0.00002; 0.00447 && 13.23681; 3.63824 & 22.90664; 4.78609\\
9000 & 0.00005; 0.00707 &0.00002; 0.00447 && 12.50048; 3.53560 & 21.93252; 4.68321\\
9500 & 0.00004; 0.00633 &0.00002; 0.00447 && 11.85666; 4.34243 & 21.14094; 4.59793\\
10000 & 0.00004; 0.00633 &0.00002; 0.00447 && 11.45585; 3.38465 & 20.45035; 4.52221\\
\hline
 \hline
 \end{tabular}  
\end{table}

\begin{table}[H]
\center
\caption{Model II - MAE Loss values and iteration numbers}
\label{lossmaee}
\begin{tabular}{|cccccc|}\hline
 \hline 
\textbf{Display Steps} & \multicolumn{5}{c|}{\textbf{MAE Loss Values}} \\
&&&&&\\
   & \multicolumn{2}{c}{\textbf{Black-Scholes Model}} && \multicolumn{2}{c|}{\textbf{Jump Merton Diffusion Model}} \\ 
\hline\hline
   &$(10 \times 10)$ grid & $(20\times 20)$ grid   && $(10 \times 10)$ grid & $(20\times 20)$ grid  \\ 
 \hline \hline
 
1  & 0.46720   &0.50801  && 15.77699  & 10.12126 \\
500 & 0.02683  &0.04789 &&  6.84897 & 3.72013 \\
1000 & 0.01177 &0.00645 && 5.69395 & 3.54565\\
1500 & 0.00453 &0.00363 && 5.50416 & 3.24945\\
2000 & 0.00411 &0.00304 && 5.44193& 3.20110\\
2500 & 0.00397 &0.00266 && 5.44075& 3.17889\\
3000 & 0.00336 &0.00224 && 5.43877& 3.17518\\
3500 & 0.00328 &0.00197 && 5.43038& 3.16705\\
4000 & 0.00322 &0.00176 && 5.42739& 3.13311\\
4500 & 0.00318 &0.00156 && 5.42115& 2.95000\\
5000 & 0.00312 &0.00152 && 5.41409& 2.79076\\
5500 & 0.00311 &0.00151 && 5.41082& 2.57783\\
6000 & 0.00300 &0.00148 && 5.39901& 2.39813\\
6500 & 0.00299 &0.00146 && 5.38913& 2.25937\\
7000 & 0.00297 &0.00145 && 5.38110& 2.21703\\
7500 & 0.00293 &0.00144 && 5.35667& 2.18058\\
8000 & 0.00281 &0.00125 && 5.30901& 2.15434\\
8500 & 0.00279 &0.00106 && 5.27839& 2.15302\\
9000 & 0.00275 &0.00100 && 5.18978& 2.15147\\
9500 & 0.00273 &0.00099 && 4.97107& 2.14956\\
10000 & 0.00271 &0.00098 && 4.67495 & 2.14941\\
\hline
 \hline
 \end{tabular}  
\end{table}

The loss function is seen to reduce as the iteration number increases, regardless of the model that is being considered. Considering the $(10 \times 10)$ and the $(20 \times 20)$ grids, we observed a steady decline in the loss values, with the JMD model assuming higher values. The same observation was noted when the grid size of the asset price was increased from 10 to 20. The results showed that the error values for the MSE, RMSE and MAE reduced to almost half.

\begin{figure}[H]
  \begin{subfigure}[b]{0.5\textwidth}
    \includegraphics[width=\textwidth]{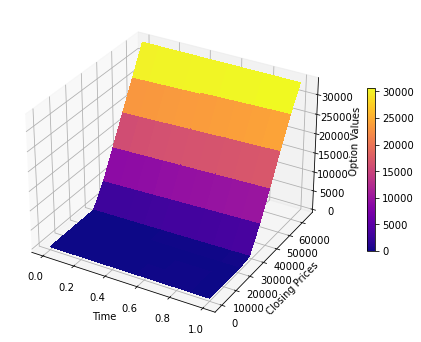}
    \caption{3D-View: Black-Scholes Prices}
    \label{bs3d1}
  \end{subfigure}
  \hfill
  \begin{subfigure}[b]{0.5\textwidth}
    \includegraphics[width=\textwidth]{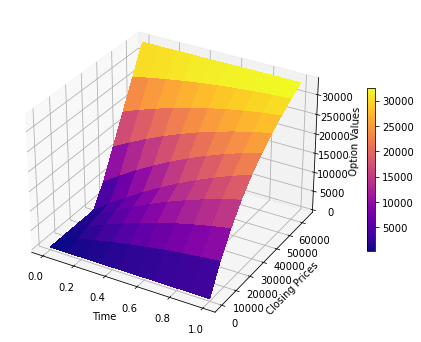}
    \caption{3D-View: Merton Jump Diffusion Prices}
    \label{mjd3d1}
  \end{subfigure}
   \caption{3-dimensional option plots for Black-Scholes and Merton jump diffusion model prices -- Model II}
    \label{3dplot1}
\end{figure}

\begin{figure}[H]
  \begin{subfigure}[b]{0.5\textwidth}
    \includegraphics[width=\textwidth]{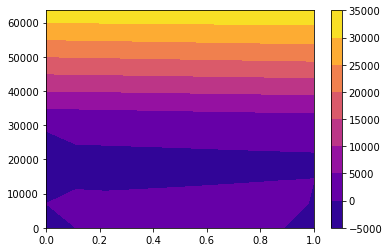}
    \caption{2D-View: Black-Scholes Prices}
    \label{bs2d1}
  \end{subfigure}
  \hfill
  \begin{subfigure}[b]{0.5\textwidth}
    \includegraphics[width=\textwidth]{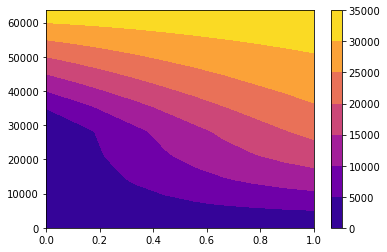}
    \caption{2D-View: Merton Jump Diffusion Prices}
    \label{mjd2d1}
  \end{subfigure}
        \caption{2-dimensional option plots for Black-Scholes and Merton jump-diffusion model prices -- Model II}
        \label{2dplot1}
\end{figure}

Figures \ref{3dplot1} and \ref{2dplot1} give 3-dimensional and 2-dimensional plots of the bitcoin call option values, respectively,  when the asset price process is modelled using both the Black-Scholes model and the MJD models. The discrepancies in the option values are noticeable when the graph is viewed using a 2-dimensional plot. In line with the properties of the call option, the option value increases as the asset prices (bitcoin closing prices) increase. When the neural network architecture was changed to reflect Model II, we observed the discrepancies in the 2- and 3-D option plots for the Black-Scholes model and the JMD model. Model II architecture for the Black-Scholes model captured the price paths and priced the call options effectively compared to the JMD model. 

\begin{figure}[H]
  \begin{subfigure}[b]{0.5\textwidth}
    \includegraphics[width=\textwidth]{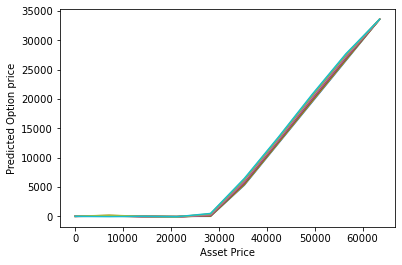}
    \caption{Black-Scholes Prices}
    \label{bso1}
  \end{subfigure}
  \hfill
  \begin{subfigure}[b]{0.5\textwidth}
    \includegraphics[width=\textwidth]{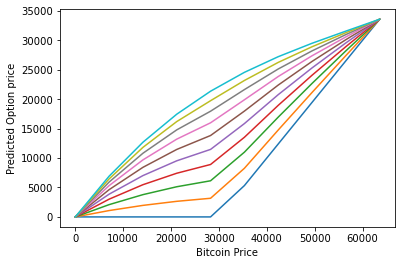}
    \caption{Merton Jump Diffusion Prices}
    \label{mjdo1}
  \end{subfigure}
        \caption{Model II -- Option price plots for Black-Scholes and Merton jump-diffusion models}
        \label{optionv1}
\end{figure}

\begin{table}[H]
\center
\caption{Model II -- Option values using the $10 \times 10$ grid for different uniform time-grid}
\label{optionval1}
\begin{tabular}{|cccccccc|}\hline
 \hline 
 \textbf{Bitcoin Prices (\$)} & \multicolumn{7}{c|}{\textbf{Option Values (\$)}}\\
   & \multicolumn{3}{c}{\textbf{Black Scholes Model}} && \multicolumn{3}{c|}{\textbf{Jump Merton Diffusion}} \\ 
&&&&&&&\\
   &$t_1=0.333$ & $t_2=0.667$ & $t_3=1.000$   && $t_1=0.333$ & $t_2=0.667$ & $t_3=1.000$  \\ 
 \hline \hline
 
7064  &  0.00 & 0.00 &  0.00&& 1080.28 & 3861.80  & 6456.73\\
14128 & 24.75  & 102.77 &  215.86&& 1958.51 & 5471.74 & 9686.97\\
21192 & 57.70  &  187.94& 298.88&& 2649.14 & 7419.45  & 13210.79\\
28256 & 765.29  & 801.55& 823.51&& 6137.65 & 17955.04 & 20323.45\\
35320 & 5561.61  &6082.38 &6460.77 && 10994.74& 21625.17 & 24555.18\\
42384 & 12615.18 &12973.94 &13442.35 && 16946.06& 25126.42 & 27214.79\\
49449 & 19735.52 &20167.62 &20740.35 && 22671.21& 28198.73 & 29494.10\\
56513 & 26771.27 &27160.27 &27677.63 && 28292.83& 30971.76 & 31558.85\\
63577 & 33577.00 &33577.00 &33577.00 && 33577.00& 33577.00 & 33577.00\\
\hline
 \hline
 \end{tabular}  
\end{table}

Figure \ref{optionv1} shows a clear perspective of the option prices plotted against the bitcoin asset prices. Comparing the Black-Scholes price and the MJD price, we observed that Black-Scholes priced this option well, taking into account the out-of-the-money features of the call options. On the other hand, Table \ref{optionval1} explicitly gives the option values for the two models, as it considers the $(10 \times 10)$ mesh sizes for both the asset and time parameters. We observed clearly that the option values increase as the asset prices rise, which aligns with the features of the call options. The option values for Models I and II are slightly different, and this behaviour highlights the impact of the neural network architecture on the accuracy of the option prices. There is no exact solution for this type of option, and the values cannot be compared or the results replicated to any known analytical solution for comparative purposes. Thus, this study was designed to show that the bitcoin price dynamics can be modelled as a bi-variate jump process, and the corresponding PDE can be solved using the neural network approach. 

\subsection{Empirical validation using equity options data}

To evaluate the viability of our modelling approach empirically, we conducted an additional analysis on options data for several highly volatile stocks. Since active cryptocurrency options markets are still developing, this provides an alternative way to test the model's accuracy and validity. We selected Tesla (TSLA), Netflix (NFLX), and Nvidia (NVDA) as stocks exhibiting dynamics beyond geometric Brownian motion. Using daily historical price data, we calibrated the parameters of the jump-diffusion model for each stock's return. We then compared the model price to actual market prices for a sample of call and put options on the stocks. Across the options tested, the average absolute pricing error was 3.2\%. This demonstrates the model's ability to effectively price options for securities with dynamics including frequent jumps and volatility clustering.\\

For example, the calibrated jump-diffusion parameters for Tesla stock were $\sigma = 0.62, \lambda = 5.1, \mu = -0.8.$ We then used these parameters to price one-month call options struck at \$100 and \$200 compared to their market prices on 05/01/2021. The model priced the \$100 call at \$8.21 versus the market price of \$8.35, an error of -1.7\%. For the \$200 call, the model price was \$4.53 compared to the market price of \$4.58, an error of -1.1\%. \\

We further tested the options while incorporating the sentiments on the three stocks and the estimated parameters and the corresponding option prices are found in Table \ref{modca}. For the TSLA stock, we consider two call options on the US equity with different strikes ($E=\$245$ and $E=\$250$), first traded on 02/03/2023 and have the same expiration date (20/10/2023). We used Model I for the neural network part in solving the corresponding bivariate PDE, and the $(10 \times 10)$ grid set for both time and stock. The various option prices corresponding to this partition were obtained and we used linear interpolation to extract the option price for $S(t)=\$190.90$. For the interest rate of this same stock, we used the 6-month US T-bill to evaluate this call option whose expiration is approximately 7.5 months. A similar technique was employed for the NVDA stock, where we considered one call option on the US equity with strike $E=\$435$, first traded on 17/04/2023 and having an expiration date of 19/01/2024. Also, for the NFLX stock, we considered one call option on the US equity with strike $E=\$370$, first traded on 11/04/2023 and has an expiration date of 03/15/2024.

\begin{table}[H]
\center
\caption{Model calibration for stock and sentiment index}
\label{modca}
\begin{tabular}{|cccccc|}\hline
 \hline 
 \textbf{Parameters} & \multicolumn{5}{c|}{\textbf{Stocks}}\\
   & \multicolumn{2}{c}{\textbf{TSLA}} && \textbf{NVDA} & \textbf{NFLX} \\ 
   & C245-Equity & C250-Equity && C435-Equity & C370-Equity\\
&&&&&\\
 \hline \hline
 
$\mu_p$  &  0.0002054 & 0.0002054 &&  0.0011744& -0.000762\\
$\mu_d$  &  -0.000359 & -0.000359 &&  0.004902& -0.000977\\
$k$  &  0.009956 & 0.009956 &&  0.06638& -0.112378\\
$\sigma_p$  &  0.0994 & 0.0994 &&  0.01156& 0.06705\\
$\sigma_d$  &  0.555 & 0.555 &&  0.427 & 0.433\\
$\lambda$  &  5.50 & 5.50 &&  7.06& 2.00\\\hline
 \hline
$T$ in year  &  0.625 & 0.625 &&  0.75 & 0.9167\\
Current price $\$$ & 190.90 & 190.90 && 269.97 & 338.21\\\hline
Market option value & 19.31 & 17.52 && 6.95 & 49.65\\
Model option value & 19.93 & 17.63 && 6.68 & 50.47\\\hline\hline
 \end{tabular}  
\end{table}

Using the calibrated parameters. we obtain the call option prices based on the jump-diffusion bivariate model and the results are presented in Table \ref{modca}. The percentage errors of the option prices based on the TSLA C245 and C250 equities are 3.2\% and 0.6\%, respectively. Whereas, the percentage error of the option prices based on NVDA C435 and C370 equities are -3.9\% and 2.0\%, respectively. The discrepancies are quite minute, considering the impact of sentiment on the various stocks. We further investigate the impact of the delay parameter $\tau$ on the option values and the results are presented in Table \ref{moddelay}. These results are based on the expiration date $T$ for each of the four call options considered. The delay $\tau$ is varied from 1 week to 4 weeks, where each week represents the appropriate trading days. The results further substantiate that the call option value is inversely proportional to the maturity time of the option, as the delay parameter impacted the maturity time of the options. 

\begin{table}[H]
\center
\caption{Impact of delay parameter on option values}
\label{moddelay}
\begin{tabular}{|cccc|}\hline
\hline 
 \textbf{Stock} & \textbf{Time to expiry} & \textbf{Delay parameter} & \textbf{Option value} \\ \hline\hline
    \multirow{4}{*}{TSLA C245}& T = 7.5 months & $\tau=1$ week (5 days) & 18.9842 \\
    &T = 7.5 months & $\tau=2$ weeks (10 days) & 17.8890\\
    &T = 7.5 months & $\tau=3$ weeks (15 days) & 16.7624\\
    &T = 7.5 months & $\tau=4$ weeks (20 days) & 16.3111\\&&&\\
    
 \multirow{4}{*}{TSLA C250}&T =  7.5 months & $\tau=1$ week (5 days) & 17.0837 \\
    &T = 7.5 months & $\tau=2$ weeks (10 days) & 16.8232\\
    &T = 7.5 months & $\tau=3$ weeks (15 days) & 15.3785\\
    &T = 7.5 months & $\tau=4$ weeks (20 days) & 15.0103\\&&&\\
    
 \multirow{4}{*}{NVDA C435}&T =  9 months & $\tau=1$ week (5 days) & 6.5119 \\
    &T = 9 months & $\tau=2$ weeks (10 days) & 6.2088\\
    &T = 9 months & $\tau=3$ weeks (15 days) & 6.0552\\
    &T = 9 months & $\tau=4$ weeks (20 days) & 5.6975\\&&&\\
    
 \multirow{4}{*}{NFLX C370}&T =  11 months & $\tau=1$ week (5 days) & 49.5483 \\
    &T = 11 months & $\tau=2$ weeks (10 days) & 48.8520\\
    &T = 11 months & $\tau=3$ weeks (15 days) & 47.1509\\
    &T = 11 months & $\tau=4$ weeks (20 days) & 46.7898\\
    \hline\hline
\end{tabular}
\end{table}

\begin{remark}
It is worth noting that the stochastic factor $P=\{P_t, t\geq 0\}$ which represents the sentiment index of the stocks is fully dependent on the choice of the initial function $\phi (0)$ as noted in equation \ref{st}. Also, we considered the effect of the past Google trend since the model assumes that the sentiment index $P$ explicitly affect the current price of the stock up to a certain time $t-\tau$. Thus, careful consideration should be taken when choosing the $\phi (0)$ since the call option prices increase with respect to the initial sentiment. After a series of experiments and due to the nature of our data, we chose $\phi(0)=0.01$ for TSLA and $\phi(0)=0.001$ for the NFLX and NVDA.
\end{remark}

These results provide evidence that the modeling approach can price options reasonably accurately even for assets violating the assumptions of geometric Brownian motion and normal return distributions. Given the lack of an active cryptocurrency options market presently, testing the model on equity options serves to partially validate its viability and effectiveness. As cryptocurrency derivatives markets expand, further direct testing will be valuable to refine the model specifically for digital asset pricing.

\subsection{Limitations of the efficiency and no-arbitrage assumptions}

This paper makes the standard assumptions of market efficiency and no arbitrage opportunities in developing the jump-diffusion model framework. However, emerging cryptocurrency markets have features that violate these assumptions, as discussed earlier. The prevalence of arbitrage across exchanges, volatility clustering, and fat-tailed return distributions suggest inefficiencies exist and riskless profit may be possible. Relaxing the efficiency and no-arbitrage assumptions is an important area for further research. Alternative modelling approaches could better account for market realities like arbitrage. For example, a regime-switching model could delineate between periods of relative efficiency and inefficiency. Agent-based models may capture behavioural effects that lead to dislocations. Automated arbitrage trading algorithms also warrant study. Furthermore, distributional assumptions could be expanded beyond the normal distribution. Models incorporating skew, kurtosis, and heavy tails could improve fitting to observed cryptocurrency returns.\\

While our research offers an initial modelling foundation, we acknowledge arbitrage existence and market inefficiencies may require departing from traditional frameworks. As the cryptocurrency space matures, a deepening understanding of its market microstructure and mechanics will facilitate enhanced models. Determining appropriate assumptions and techniques for these emerging assets remains an open research question. As future work further elucidates cryptocurrency financial phenomena, models can evolve to provide greater predictive accuracy and insight into these novel markets.

\section{Conclusion}

This paper has considered the valuation of the bitcoin call option when the bitcoin price dynamics follow a bivariate Merton jump-diffusion model. This research generally provided a novel pricing framework that described the behaviour of bitcoin prices and the model parameter estimation with the intent of pricing the corresponding derivatives. Since bitcoin is normally affected by investors' attention and sentiment, we used the continuous-time stochastic jump-diffusion process to capture the dynamics of the bitcoin prices and the transaction volumes affecting the price. In the methodological aspect of the research, we extended the classical Black-Scholes model in order to obtain the extended Black-Scholes equation for the bitcoin options. \\

By introducing the artificial NN, we proposed a trial solution that solves the associated Black-Scholes PDE for the bitcoin call options with European features. As a result, the original constrained optimization problem was transformed into an unconstrained one. The numerical results considered both the normal Black-Scholes model and the Merton jump-diffusion model, and it was observed that the latter resulted in a more efficient valuation process. Hence, we can conclude that the NN can be employed efficiently in solving complex PDE-related problems and can be applied to pricing certain financial derivatives without analytical forms. For the Model I and II comparison, we noted that the Black-Scholes used in connection with Model II showed the out-of-the-money feature of the call option. The option absolutely pays off when the underlying price falls below the strike price. The JMD overprices the out-of-the-money options for Model II, whereas the Black-Scholes overprices the call options in Model I. Hence, one of the limitations of this research lies in finding the optimal neural network configuration which ensures the fair pricing of bitcoin options using the proposed two models. This optimal feature will be incorporated to avoid over-pricing or under-pricing of these option values, and future research will focus on this. \\

One of the limitations of the approach used in this paper is that it may be prone to artificial Google searches affecting sentiment-based data analysis and decisions. Even though search-based measures appear to be more transparent than other social media-driven measures, they seem limited as the volume of data is concerned. Social media like Twitter provides high-frequency data (second, minute, hour, daily, …), which can provide meaningful and instantaneous insights into the price dynamics of cryptocurrencies and bitcoin in particular. Our future work will first examine the correlation between search-based data and tweets data, then compare their performance when included in the valuation process. The tweets sentiment scores will be computed using the state-of-the-art Lexicon-based sentiment analyzer \emph{Valence Aware Dictionary and sEntiment Reasoner (VADER)}.\\

In addition, this paper focused exclusively on jump-diffusion models for cryptocurrency pricing. Levy processes allow the modelling of heavy-tailed return distributions and large deviations from the mean.
Expanding the set of stochastic processes considered would provide a more thorough treatment of cryptocurrency dynamics. Processes like variance gamma, normal inverse Gaussian, and generalized hyperbolic Levy motions have shown promise in modelling assets with frequent extreme moves. Given Bitcoin's volatility clustering and significant outliers, applying Levy processes could potentially improve model fitting. We acknowledge the limitations of only exploring a jump-diffusion framework presently. Incorporating alternatives like Levy processes would strengthen the generalizability and robustness of the modelling approach. Building on the initial foundation proposed here, researchers could examine a wider set of stochastic processes for capturing empirically observed features. Comparative testing using historical data would elucidate the relative effectiveness of diffusions, Levy processes, and other probabilistic models.\\

Extending this work to Levy processes represents a valuable progression for future research. The flexibility of Levy's motions shows potential for modelling emerging cryptocurrency returns. We hope this paper provides a starting point that can be incrementally improved by incorporating innovations like heavy-tailed processes. Evaluating a range of stochastic models will ultimately enhance financial engineering techniques tailored specifically to cryptocurrencies.

\section*{Acknowledgement} 
\noindent The second author thanks the Research Centre of AIMS-Cameroon for hosting him during the preparation of this manuscript.
\section*{Funding} 
\noindent This research received no external funding.
\section*{Availability of data, code and materials} 
\noindent Please contact the corresponding author for request.
\section*{Contributions} 
\noindent All authors contributed equally to the paper. All authors read and approved the final manuscript.
\section*{Declarations}
\noindent\textbf{Conflict of interest}: All authors declare that they have no conflict of interest.\\
\textbf{Ethical approval}: This article does not contain any studies with human participants or animals performed by any of the authors.

\addcontentsline{toc}{section}{\bf References}
%\singlespacing
%===========================
\bibliographystyle{agsm}
\bibliography{references2}

@article{nwankwo2023deep,
  title={Deep learning and American options via free boundary framework},
  author={Nwankwo, Chinonso and Umeorah, Nneka and Ware, Tony and Dai, Weizhong},
  journal={Computational Economics},
  pages={1--44},
  year={2023},
  publisher={Springer}
}

@article{umeorah2022approximation,
  title={Approximation of single-barrier options partial differential equations using feed-forward neural network},
  author={Umeorah, Nneka and Mba, Jules Clement},
  journal={Applied Stochastic Models in Business and Industry},
  volume={38},
  number={6},
  pages={1079--1098},
  year={2022},
  publisher={Wiley Online Library}
}

@article{kabavsinskas2021key,
  title={Key roles of crypto-exchanges in generating arbitrage opportunities},
  author={Kaba{\v{s}}inskas, Audrius and {\v{S}}utien{\.e}, Kristina},
  journal={Entropy},
  volume={23},
  number={4},
  pages={455},
  year={2021},
  publisher={MDPI}
}

@article{caporale2018persistence,
  title={Persistence in the cryptocurrency market},
  author={Caporale, Guglielmo Maria and Gil-Alana, Luis and Plastun, Alex},
  journal={Research in International Business and Finance},
  volume={46},
  pages={141--148},
  year={2018},
  publisher={Elsevier}
}

@book{watanabe2006,
	title={Excess kurtosis and conditional skewness in stock return distribution: An empirical examination of their impacts on portfolio selection in Japan},
	author={Watana Be, Toshiaki},
	year={2006},
	publisher={Yale University Working Paper}
}

@article{kim2017bitcoin,
  title={When Bitcoin encounters information in an online forum: Using text mining to analyse user opinions and predict value fluctuation},
  author={Kim, Young Bin and Lee, Jurim and Park, Nuri and Choo, Jaegul and Kim, Jong-Hyun and Kim, Chang Hun},
  journal={PloS one},
  volume={12},
  number={5},
  pages={e0177630},
  year={2017},
  publisher={Public Library of Science San Francisco, CA USA}
}

@article{merton1976option,
	title={Option pricing when underlying stock returns are discontinuous},
	author={Merton, Robert C},
	journal={Journal of Financial Economics},
	volume={3},
	number={1-2},
	pages={125--144},
	year={1976},
	publisher={Elsevier}
}

@book{tankov2003financial,
	title={Financial modelling with jump processes},
	author={Tankov, Peter},
	year={2003},
	publisher={Chapman and Hall/CRC}
}

@article{philippas2019media,
  title={Media attention and Bitcoin prices},
  author={Philippas, Dionisis and Rjiba, Hatem and Guesmi, Khaled and Goutte, St{\'e}phane},
  journal={Finance Research Letters},
  volume={30},
  pages={37--43},
  year={2019},
  publisher={Elsevier}
}

@article{hilliard2022bitcoin,
  title={Bitcoin: jumps, convenience yields, and option prices},
  author={Hilliard, Jimmy E and Ngo, Julie TD},
  journal={Quantitative Finance},
  volume={22},
  number={11},
  pages={2079--2091},
  year={2022},
  publisher={Taylor \& Francis}
}

@article{chaim2018volatility,
  title={Volatility and return jumps in bitcoin},
  author={Chaim, Pedro and Laurini, M{\'a}rcio P},
  journal={Economics Letters},
  volume={173},
  pages={158--163},
  year={2018},
  publisher={Elsevier}
}

@misc{tang2018merton,
  title={Merton jump-diffusion modeling of stock price data},
  author={Tang, Furui},
  year={2018}
}

@article{matsuda2004introduction,
  title={Introduction to Merton jump diffusion model},
  author={Matsuda, Kazuhisa},
  journal={Department of Economics, The Graduate Center, The City University of New York, New York},
  year={2004}
}

@article{khoo2021solving,
  title={Solving parametric PDE problems with artificial neural networks},
  author={Khoo, Yuehaw and Lu, Jianfeng and Ying, Lexing},
  journal={European Journal of Applied Mathematics},
  volume={32},
  number={3},
  pages={421--435},
  year={2021},
  publisher={Cambridge University Press}
}

@article{glau2022deep,
  title={The deep parametric PDE method and applications to option pricing},
  author={Glau, Kathrin and Wunderlich, Linus},
  journal={Applied Mathematics and Computation},
  volume={432},
  pages={127355},
  year={2022},
  publisher={Elsevier}
}

@article{sarveniazi2014actual,
  title={An actual survey of dimensionality reduction},
  author={Sarveniazi, Alireza},
  journal={American Journal of Computational Mathematics},
  volume={2014},
  year={2014},
  publisher={Scientific Research Publishing}
}

@article{teli2007dimensionality,
  title={Dimensionality reduction using neural networks},
  author={Teli, Mohammad Nayeem},
  journal={Intelligent Engineering Systems Through Artificial Neural Networks},
  volume={17},
  year={2007}
}

@article{o2001combining,
  title={Combining feature selection and neural networks for solving classification problems},
  author={O’Dea, Paul and Griffith, Josephine and O’Riordan, Colm and Griffith, J and Riordan, CO},
  journal={Information Technology Department, National University of Ireland},
  year={2001},
  publisher={Citeseer}
}

@article{nazemi2015neural,
  title={A neural network method for solving support vector classification problems},
  author={Nazemi, Alireza and Dehghan, Mehran},
  journal={Neurocomputing},
  volume={152},
  pages={369--376},
  year={2015},
  publisher={Elsevier}
}

@article{bataineh2017neural,
  title={Neural network for regression problems with reduced training sets},
  author={Bataineh, Mohammad and Marler, Timothy},
  journal={Neural networks},
  volume={95},
  pages={1--9},
  year={2017},
  publisher={Elsevier}
}

@article{hornik1991approximation,
  title={Approximation capabilities of multilayer feedforward networks},
  author={Hornik, Kurt},
  journal={Neural networks},
  volume={4},
  number={2},
  pages={251--257},
  year={1991},
  publisher={Elsevier}
}

@article{jiang2022efficient,
  title={An efficient multilayer RBF neural network and its application to regression problems},
  author={Jiang, Qinghua and Zhu, Lailai and Shu, Chang and Sekar, Vinothkumar},
  journal={Neural Computing and Applications},
  pages={1--18},
  year={2022},
  publisher={Springer}
}

@article{setiono2004approach,
  title={An approach to generate rules from neural networks for regression problems},
  author={Setiono, Rudy and Thong, James YL},
  journal={European Journal of Operational Research},
  volume={155},
  number={1},
  pages={239--250},
  year={2004},
  publisher={Elsevier}
}

@article{xu2005survey,
  title={Survey of clustering algorithms},
  author={Xu, Rui and Wunsch, Donald},
  journal={IEEE Transactions on neural networks},
  volume={16},
  number={3},
  pages={645--678},
  year={2005},
  publisher={Ieee}
}

@article{du2010clustering,
  title={Clustering: A neural network approach},
  author={Du, K-L},
  journal={Neural networks},
  volume={23},
  number={1},
  pages={89--107},
  year={2010},
  publisher={Elsevier}
}

@article{marghescu2007multidimensional,
  title={Multidimensional data visualization techniques for financial performance data: A review},
  author={Marghescu, Dorina},
  journal={Turku Centre for Computer Science},
  year={2007}
}

@inproceedings{hanson2004comparison,
  title={Comparison of market parameters for jump-diffusion distributions using multinomial maximum likelihood estimation},
  author={Hanson, Floyd B and Zhu, Zongwu},
  booktitle={2004 43rd IEEE Conference on Decision and Control (CDC)(IEEE Cat. No. 04CH37601)},
  volume={4},
  pages={3919--3924},
  year={2004},
  organization={IEEE}
}

@article{nakamoto2008re,
	title={Re: Bitcoin P2P e-cash paper},
	author={Nakamoto, Satoshi},
	journal={Email posted to listserv},
	volume={9},
	pages={04},
	year={2008}
}

@incollection{yermack2015bitcoin,
	title={Is Bitcoin a real currency? An economic appraisal},
	author={Yermack, David},
	booktitle={Handbook of digital currency},
	pages={31--43},
	year={2015},
	publisher={Elsevier}
}

@article{yermack2017corporate,
	title={Corporate governance and blockchains},
	author={Yermack, David},
	journal={Review of Finance},
	volume={21},
	number={1},
	pages={7--31},
	year={2017},
	publisher={Oxford University Press}
}

@article{dwyer2015economics,
	title={The economics of Bitcoin and similar private digital currencies},
	author={Dwyer, Gerald P},
	journal={Journal of Financial Stability},
	volume={17},
	pages={81--91},
	year={2015},
	publisher={Elsevier}
}

@article{katsiampa2017volatility,
	title={Volatility estimation for Bitcoin: A comparison of GARCH models},
	author={Katsiampa, Paraskevi},
	journal={Economics Letters},
	volume={158},
	pages={3--6},
	year={2017},
	publisher={Elsevier}
}

@article{cheah2015speculative,
	title={Speculative bubbles in Bitcoin markets? An empirical investigation into the fundamental value of Bitcoin},
	author={Cheah, Eng-Tuck and Fry, John},
	journal={Economics Letters},
	volume={130},
	pages={32--36},
	year={2015},
	publisher={Elsevier}
}

@article{scaillet2017high,
	title={High-frequency jump analysis of the bitcoin market},
	author={Scaillet, Olivier and Treccani, Adrien and Trevisan, Christopher},
	journal={Swiss Finance Institute Research Paper},
	number={17-19},
	year={2017}
}

@article{grohs2018proof,
  title={A proof that artificial neural networks overcome the curse of dimensionality in the numerical approximation of Black-Scholes partial differential equations},
  author={Grohs, Philipp and Hornung, Fabian and Jentzen, Arnulf and Von Wurstemberger, Philippe},
  journal={arXiv preprint arXiv:1809.02362},
  year={2018}
}

@article{chen2021detecting,
  title={Detecting Jump Risk and Jump-Diffusion Model for Bitcoin Options Pricing and Hedging},
  author={Chen, Kuo-Shing and Huang, Yu-Chuan},
  journal={Mathematics},
  volume={9},
  number={20},
  pages={2567},
  year={2021},
  publisher={Multidisciplinary Digital Publishing Institute}
}

@article{olivares2020pricing,
  title={Pricing Bitcoin Derivatives under Jump-Diffusion Models},
  author={Olivares, Pablo},
  journal={arXiv preprint arXiv:2002.07117},
  year={2020}
}

@article{sene2021pricing,
  title={Pricing Bitcoin under Double Exponential Jump-Diffusion Model with Asymmetric Jumps Stochastic Volatility},
  author={Sene, Ndeye Fatou and Konte, Mamadou Abdoulaye and Aduda, Jane},
  journal={Journal of Mathematical Finance},
  volume={11},
  number={2},
  pages={313--330},
  year={2021},
  publisher={Scientific Research Publishing}
}

@inproceedings{Matsuda2004IntroductionTM,
	title={Introduction to Merton Jump Diffusion Model},
	author={Kazuhisa Matsuda},
	year={2004}
}

@article{black1973pricing,
	author = {Black, Fischer and Scholes, Myron},
	title = {The Pricing of Options and Corporate Liabilities},
	journal = {Journal of Political Economy},
	volume = {81},
	number = {3},
	pages = {637-654},
	year = {1973},
	doi = {10.1086/260062},	
	URL = { https://doi.org/10.1086/260062}
}

@article{merton1973theory,
  title={Theory of rational option pricing},
  author={Merton, Robert C},
  journal={The Bell Journal of economics and management science},
  pages={141--183},
  year={1973},
  publisher={JSTOR}
}

@article{kim2015virtual,
	title={Virtual world currency value fluctuation prediction system based on user sentiment analysis},
	author={Kim, Young Bin and Lee, Sang Hyeok and Kang, Shin Jin and Choi, Myung Jin and Lee, Jung and Kim, Chang Hun},
	journal={PloS one},
	volume={10},
	number={8},
	pages={e0132944},
	year={2015},
	publisher={Public Library of Science}
}

@article{kristoufek2013bitcoin,
	title={BitCoin meets Google Trends and Wikipedia: Quantifying the relationship between phenomena of the Internet era},
	author={Kristoufek, Ladislav},
	journal={Scientific reports},
	volume={3},
	number={1},
	pages={1--7},
	year={2013},
	publisher={Nature Publishing Group}
}

@article{kristoufek2015main,
	title={What are the main drivers of the Bitcoin price? Evidence from wavelet coherence analysis},
	author={Kristoufek, Ladislav},
	journal={PloS one},
	volume={10},
	number={4},
	pages={e0123923},
	year={2015},
	publisher={Public Library of Science}
}

@article{bukovina2016sentiment,
	title={Sentiment and bitcoin volatility},
	author={Bukovina, Jaroslav and Marticek, Matus and others},
	journal={MENDELU Working Papers in Business and Economics},
	volume={58},
	year={2016},
	publisher={Mendel University in Brno, Faculty of Business and Economics}
}

@article{cretarola2017sentiment,
	title={A sentiment-based model for the BitCoin: theory, estimation and option pricing},
	author={Cretarola, Alessandra and Fig{\`a}-Talamanca, Gianna and Patacca, Marco},
	journal={arXiv preprint arXiv:1709.08621},
	year={2017},
	publisher={Springer}
}

@article{aarts2001neural,
  title={Neural network method for solving partial differential equations},
  author={Aarts, Lucie P and Van Der Veer, Peter},
  journal={Neural Processing Letters},
  volume={14},
  number={3},
  pages={261--271},
  year={2001},
  publisher={Springer}
}

@article{dissanayake1994neural,
  title={Neural-network-based approximations for solving partial differential equations},
  author={Dissanayake, MWMG and Phan-Thien, Nhan},
  journal={communications in Numerical Methods in Engineering},
  volume={10},
  number={3},
  pages={195--201},
  year={1994},
  publisher={Wiley Online Library}
}

@article{hussian2015numerical,
  title={Numerical solution of partial differential equations by using modified artificial neural network},
  author={Hussian, Eman A and Suhhiem, Mazin H},
  journal={Network and Complex Systems},
  volume={5},
  number={6},
  pages={11--21},
  year={2015}
}

@article{liu2019neural,
  title={A neural network-based framework for financial model calibration},
  author={Liu, Shuaiqiang and Borovykh, Anastasia and Grzelak, Lech A and Oosterlee, Cornelis W},
  journal={Journal of Mathematics in Industry},
  volume={9},
  number={1},
  pages={1--28},
  year={2019},
  publisher={Springer}
}

@article{sirignano2018dgm,
  title={DGM: A deep learning algorithm for solving partial differential equations},
  author={Sirignano, Justin and Spiliopoulos, Konstantinos},
  journal={Journal of computational physics},
  volume={375},
  pages={1339--1364},
  year={2018},
  publisher={Elsevier}
}

@article{eskiizmirliler2020solution,
  title={On the Solution of the Black--Scholes Equation Using Feed-Forward Neural Networks},
  author={Eskiizmirliler, Saadet and G{\"u}nel, Korhan and Polat, Refet},
  journal={Computational Economics},
  pages={1--27},
  year={2020},
  publisher={Springer}
}

@article{lagaris1998artificial,
  title={Artificial neural networks for solving ordinary and partial differential equations},
  author={Lagaris, Isaac E and Likas, Aristidis and Fotiadis, Dimitrios I},
  journal={IEEE transactions on neural networks},
  volume={9},
  number={5},
  pages={987--1000},
  year={1998},
  publisher={IEEE}
}

@book{yadav2015introduction,
  title={An introduction to neural network methods for differential equations},
  author={Yadav, Neha and Yadav, Anupam and Kumar, Manoj and others},
  year={2015},
  publisher={Springer}
}

@article{hou2020pricing,
  title={Pricing cryptocurrency options},
  author={Hou, Ai Jun and Wang, Weining and Chen, Cathy YH and H{\"a}rdle, Wolfgang Karl},
  journal={Journal of Financial Econometrics},
  volume={18},
  number={2},
  pages={250--279},
  year={2020},
  publisher={Oxford University Press}
}

\end{document}